\documentclass{article}

\usepackage{arxiv}
\usepackage{authblk}
\usepackage[sort,numbers]{natbib}
\usepackage[utf8]{inputenc} 
\usepackage[T1]{fontenc}    
\usepackage{url}            
\usepackage{hyperref}       
\usepackage{booktabs}       
\usepackage{amsfonts}       
\usepackage{amsmath,amsthm,amssymb}        
\DeclareMathOperator*{\argmax}{argmax} 
\usepackage{nicefrac}       
\usepackage{microtype}      
\usepackage{graphicx}
\usepackage{subfigure}
\usepackage{caption}
\usepackage{tablefootnote}
\usepackage[linesnumbered,ruled]{algorithm2e}

\usepackage{xcolor}
\newcommand{\R}{{\mathbb{R}}} 
\newcommand{\X}{{\mathcal{X}}} 
\newcommand{\A}{\mathcal{A}}
\newcommand{\PP}{\mathbb{P}}
\newcommand{\E}{\mathbb{E}}
\newcommand{\T}{\mathcal{T}}

\theoremstyle{definition}
\newtheorem{defn}{Definition}
\theoremstyle{plain}

\title{Evaluation of Deep Reinforcement Learning Algorithms for Portfolio Optimisation}

\author[ ]{Lu Chung I\textsuperscript{1}\thanks{Corresponding author: \texttt{lu.chung.i@u.nus.edu}}}
\affil[1]{National University of Singapore, 21 Lower Kent Ridge Road, Singapore 119077, \protect\\ Asian Institute of Digital Finance (AIDF)}

\begin{document}

\maketitle

\begin{abstract}
We evaluate benchmark deep reinforcement learning algorithms on the task of portfolio optimisation using simulated data.
The simulator to generate the data is based on correlated geometric Brownian motion with the Bertsimas-Lo market impact model.
Using the Kelly criterion (log utility) as the objective, we can analytically derive the optimal policy without market impact as an upper bound to measure performance when including market impact.
We find that the off-policy algorithms DDPG, TD3 and SAC are unable to learn the right $Q$-function due to the noisy rewards and therefore perform poorly.
The on-policy algorithms PPO and A2C, with the use of generalised advantage estimation, are able to deal with the noise and derive a close to optimal policy.
The clipping variant of PPO was found to be important in preventing the policy from deviating from the optimal once converged.
In a more challenging environment where we have regime changes in the GBM parameters, we find that PPO, combined with a hidden Markov model to learn and predict the regime context, is able to learn different policies adapted to each regime.
Overall, we find that the sample complexity of these algorithms is too high for applications using real data, requiring more than 2m steps to learn a good policy in the simplest setting, which is equivalent to almost 8,000 years of daily prices.
\end{abstract}

\section{Introduction}

The aim of this paper is to test popular benchmark deep reinforcement learning (DRL) algorithms in the task of portfolio optimisation using simulated data.
An investment universe of three correlated stocks and a cash account paying interest is considered.
We discretise a finite investment horizon into equally spaced time periods.
The agent produces portfolio weights at the start of each period and the goal is to maximise the utility of the final portfolio value.
Our choice for the utility function is the log utility, also known as the Kelly criterion for reasons discussed in Section~\ref{sec:PortfolioProblem}.

We would like to understand which of these DRL algorithms are more robust to the noisy states and rewards.
Specifically, are there any specific components of the algorithms that are critical to the performance?
Another aim of the paper is to understand the sample complexity of these algorithms applied to this task.
As we only see one single realisation of the financial markets, there is no way to reset the environment and start again.
Therefore, it is important for an algorithm to learn quickly and generalise well.

\section{Related Work}
Although the earliest works using DRL for portfolio optimisation or trading was done more than 20 years ago (see \cite{moody1998reinforcement, moody1998performance, moody2001learning}), the field has only recently gained traction.
Most of the recent work focuses on neural architecture innovations and feature selection or engineering (see e.g. \cite{jiang2017deep, xiong2018practical, ye2020reinforcement, wang2019alphastock, deng2017deep, zhang2020deep,xu2021relation}).
Practical data collection limitations, where typically only financial time series are available (and not the actions of market participants), often lead to the assumption of zero market impact.
If the state only includes the time series of the assets, this assumption means the state transition is independent of the action taken.
The problem can be reduced to a one-state MDP or equivalently a contextual multi-armed bandit problem such as in \cite{shen2015portfolio}.
We can then optimise each action in isolation instead of optimising them jointly.

However, if we include transaction cost into the problem, it creates a temporal impact of current actions on future returns of the portfolio, even though the financial time series remains independent of the action.
In such a scenario, another way to frame the problem is to define the state as the portfolio wealth and weights in the stocks or simply the dollar value in each stock such as in \cite{dixon2020g}.
The financial time series then becomes additional context and the state transition is dependent on the current state and action plus the added context.

The model used to simulate the market, described in Section~\ref{sec:Simulator}, will include market impact which also indirectly causes transaction costs.
In other words, the action will affect the financial time series directly.
The state will include both the portfolio wealth, weights in the stocks, and the financial time series.

One aspect of the work in this area is the lack of a standard benchmark, which is inherently difficult to set up.
Evaluation in the work referenced above typically selects different sets of financial time series and a set of return measures, some risk-adjusted, and does a relative comparision against baselines.
As financial markets are recognised to be highly volatile and non-stationary in nature, it is difficult to assess whether an algorithm is truly performing well and able to generalise consistently.
In other words, it is difficult to gauge whether the performance is due to luck or ability.
The choice of simulator allows us to derive the optimal policy analytically when assuming zero market impact.
When including market impact, the derived optimal policy without market impact serves as an upper bound to measure performance.

\section{Preliminaries} \label{sec:Preliminaries}

In this section, we will describe the simulator used to generate the data and the MDP framework used to model the portfolio optimisation problem.

\subsection{Simulator} \label{sec:Simulator}

The simulator has three components:
\begin{enumerate}
    \item A geometric Brownian motion (GBM) simulator to generate the price series of the stocks.
    \item A Bertsimas-Lo market impact model to simulate the impact of trades on the price series.
    \item A regime switching model to simulate the change in parameters of the GBM simulator.
\end{enumerate}
The first two components are used in the first stage of the experiments to evaluate the performance of the DRL algorithms.
The third component is used in the second stage of the experiments to evaluate the performance of the DRL algorithms in a more challenging environment.

\subsubsection{Geometric Brownian Motion} \label{sec:GBM}

The simulator generates time series of prices for the stocks using correlated geometric Brownian motion (GBM) \cite[Chapter 5.4.2]{shreve2004stochastic}.
\begin{defn}[Correlated geometric Brownian motion] \label{def:GBM}
    Let $S_i(t)$ be the price of the $i$-th stock at time $t$, $\mu_i$ be the drift, $\sigma_i$ be the volatility and $B_i(t)$ is a standard Brownian motion with $dB_idB_j=\rho_{ij}dt$ for $i\neq j$ representing the instantaneous correlation between the $i$-th and $j$th stocks for some $\rho_{ij}\in[-1,1]$.
    The dynamics of the price process $S_i(t)$ is given by the following stochastic differential equation (SDE) for $i=1,\ldots,n$ where $n$ is the number of stocks:
    \begin{equation}
        dS_i(t)=\mu_i S_i(t)dt+\sigma_i S_i(t)dB_i(t) \quad\text{for } i=1,\ldots,n
    \end{equation}
\end{defn}
We will use $n=3$ for the experiments.
Using Ito's Lemma \cite[Theorem 4.4.6]{shreve2004stochastic}, we can show that the log returns of the stocks follow a multivariate normal distribution under GBM:
\begin{equation}
    d\log{S_i(t)}=(\mu_i-\frac{\sigma_i^{2}}{2})dt+\sigma_i dB_i(t)
\end{equation}
While GBM does not reflect some important aspects of stock behaviour, such as the presence of fat tails, it captures the noisy aspect of stock prices and is the basis behind the celebrated Black-Scholes model.
This SDE has an explicit solution which makes it straightforward to simulate the price series.
Let $S_i(t)$ be the price of the $i$-th stock at time $t$ and $\Delta t$ be the time step.
Then the price at time $t+\Delta t$ is given by:
\begin{equation}
    S_i(t+\Delta t)=S_i(t)e^{(\mu_i-\frac{\sigma_i^{2}}{2})\Delta t+\sigma_i \Delta B_i}
\end{equation}
where $\Delta B_i\sim N(0,\Delta t)$ and Cov$(\Delta B_i,\Delta B_j)=\rho_{ij}\Delta t$ for $i\neq j$.

The drift, volatility and correlation are the signals that determine the optimal portfolio, which is further discussed in Section~\ref{sec:OptimalPolicy}.
The Brownian motion is the source of noise in the price series and the reason why the same action will not always produce the same result.

The drift, volatility and correlation are also the parameters to specify the simulator.
These were estimated from historical data using 3 exchange traded funds (ETFs) with tickers VUG, VTV and GLD as shown in Appendix~\ref{app:low} by calculating the mean, standard deviation and correlation of the daily log returns.
The time step $\Delta t = 1$ represents one year while each trading day is represented by a time step of $\Delta t = \frac{1}{252}$.
The ETFs are chosen as representations of portfolio allocation components with VUG being a growth stocks ETF, VTV being a value stocks ETF and GLD being a gold ETF.
The data was obtained from Yahoo Finance from 2018 to 2023.
All prices at the start of an episode are normalised to 1.

\subsubsection{Bertsimas-Lo Market Impact Model} \label{sec:BLM}

A GBM simulator alone lacks two important aspects of the real world: transaction cost and market impact.
Including transaction cost would cause actions on the portfolio to have temporal impact on future returns of the portfolio which are a key component of any reward function relating to portfolio optimisation.
Increasing the number of shares in a stock would not only incur an immediate cost but also potential future costs when liquidating the stock.
Without market impact, any action on the portfolio would not affect the stock prices which changes the nature of the problem as it becomes possible to reduce it to single period optimisation.
Therefore, we will use the Bertsimas-Lo (BL) model \cite{bertsimas1998optimal}, which uses GBM as the base price process, to model the impact of a trade on the price of the asset.
\begin{defn}[Bertsimas-Lo model] \label{def:BL}
    Let $y_t$ be the number of shares held at time $t$ and $y_0$ be the initial number of shares.
    Let $S(t)$ be the price of the asset at time $t$ following Definition \ref{def:GBM} and $\hat{S}(t)$ be the price of the asset after a trade of $y_t-y_0$ shares at time $t$.
    Then the Bertsimas-Lo (BL) model is given by:
    \begin{equation}
        \hat{S}(t) = S(t)e^{\eta(\frac{dy}{dt}) + \gamma (y_t-y_0)}
    \end{equation}
    where $\eta$ and $\gamma$ are hyperparameters that control the temporary and permanent impact respectively and $\frac{dy}{dt}$ is the trading rate.
\end{defn}
To be clear, if $y_t-y_0 > 0$, then $y_t-y_0$ is the number of shares bought and if $y_t-y_0 < 0$, then $y_t-y_0$ is the number of shares sold.
Let $Y$ be the total number of shares traded in a period and $\Delta T$ be the time duration of the transaction.
For simplicity, we will assume a constant trading rate, i.e., $\frac{dy}{dt} = \frac{Y}{\Delta T}$.
Two types of impact are included in the BL model:
\begin{enumerate}
  \item \textbf{Temporary impact} causes prices to move up for buy trades and move down for sell trades but the effect is temporary and only serves to increase the transactional cost of executing the trade.
  \item \textbf{Permanent impact} as the name suggests would cause a permanent change in the price and also increase the transactional cost.
\end{enumerate}
Both types of impact are a function of the number of shares transacted and can be tuned by hyperparameters $\eta$ and $\gamma$ respectively.

To calculate the cost of a trade, we need to calculate the integral:
\begin{equation}
  C:=\int_{t}^{t+\Delta t}{\hat{S}(t)\frac{dy}{dt}dt}
\end{equation}
where $t$ is the time of the trade.
An accurate approximation would require us to simulate intra-period prices, which is computationally expensive.
Therefore, we will do a crude approximation by assuming the unaffected stock price moves linearly from $S(t)$ to $S(t+\Delta t)$:
\begin{equation}
C\approx Y[\frac{1}{2}(1+\frac{\eta}{\Delta t}Y)(S(t+\Delta t)-S(t))+\gamma Y(\frac{1}{3}S(t+\Delta t)+\frac{1}{6}S(t))]
\end{equation}
where $C$ is the cost of the trade.
Although this reduces the noise in the cost arising from market impact, it does not change the necessity of learning a policy to optimise the impact and cost.

\subsubsection{Regime Switching Model} \label{sec:RSM}

There is evidence that financial markets exhibit dual regime behaviour \cite{ang2002international, schaller1997regime}.
Therefore, in the second stage, we turn to a regime switching simulator.
In essence, the simulator will switch between two sets of GBM parameters based on a continuous time Markov chain (CTMC).
We keep parameters of the BL model constant between the two regimes.

We discretise the CTMC \cite{doytchinov2010time} into the same periodic points used for the rebalancing of the portfolio.
The hidden regime variable $Z_t$ at time $t$ can take on a fixed number of values $z_1,z_2,\ldots,z_K$.
The transition probabilities from time $t$ to time $t+\Delta t$ are denoted by $p_{ij}(\Delta t) = \PP(Z_{t+\Delta t}=z_j|Z_t=z_i)$ which are stationary.
The set of transition probabilities is represented by a matrix $P(\Delta t)$.

The Chapman-Kolmogorov equation gives us:
\begin{equation}
    P(t + \Delta t) = P(t)P(\Delta t)
\end{equation}

We can then construct the derivative of $P(t)$:
\begin{equation}
    \begin{split}
        \frac{d}{dt}P(t)&=\lim_{\Delta t \to 0}\frac{P(t+\Delta t) - P(t)}{\Delta t}
        \\
        &=\lim_{\Delta t \to 0}\frac{P(t)P(\Delta t) - P(t)}{\Delta t}
        \\
        &=\lim_{\Delta t \to 0}\frac{P(t)}{\Delta t}(P(\Delta t)-I)
        \\
        &=P(t)Q
    \end{split}
\end{equation}
where $I$ is the identity matrix of size K and we have defined the infinitesimal generator of the CTMC as:
\begin{equation}
    Q := \lim_{\Delta t \to 0}\frac{1}{\Delta t}(P(\Delta t)-I)
\end{equation}

Finally, we can solve for $P(t)$:
\begin{equation} \label{eq:ctmc_rescale}
    P(t) = e^{Qt}
\end{equation}
We can then rescale any transition probabilities measured on different time scales using Equation \eqref{eq:ctmc_rescale}.
Following \cite{ang2002international}, we use a two-regime model based on parameters in the experiments, i.e., $K=2$.

\subsection{Markov Decision Process} \label{sec:MDP}

We frame the portfolio optimisation problem as a finite horizon Markov decision process (MDP) \cite{sutton2018reinforcement} which is defined by a tuple $(\X,\A,\PP,R,\alpha)$.
\begin{itemize}
    \item $\X \subseteq \R^m$: A set of possible states $x$ the agent can be in where $m$ is the dimension of the state space.
    \item $\A \subseteq \R^n$: A set of possible actions $a$ the agent can take where $n$ is the number of stocks.
    \item $\PP(x' | x, a)$: The state transition probability function, defining the probability of transitioning to state $x'$ after taking action $a$ in state $x$.
    \item $R: \X \times \A \times \X \ni (x, a, x') \mapsto r \in \R$: The reward function, specifying the immediate reward $r$ received after taking action $a$ in state $x$ and transitioning to state $x'$.
    \item $\alpha \in (0,1]$: The discount factor, which determines the present value of future rewards.
\end{itemize}

In the following, the investment horizon is a fixed duration $T$ and is divided into equally spaced time periods each of duration $\Delta t$.
Therefore, there are $N=\frac{T}{\Delta t}$ time periods.
The states $x_t \in \X$ and actions $a_t \in \A$ are indexed by $t=0,1,\ldots,N$ for times $0, t+\Delta t, \ldots, T$ respectively.
Similarly, we will index the $i$-th stock $S_{i,t}$ to indicate the time step.

Next, we define a value function $V_{\pi}$ as the expected sum of discounted rewards, sometimes referred to as the return, starting from state $x$ and following the policy $\pi$:
\begin{equation} \label{eq:value_function}
        V_{\pi}(x) := \E \left[ \sum_{t=0}^{N-1} \alpha^{t} R(x_t,a_t,x_{t+1}) \ \Bigg| \ x_0 = x \right]
\end{equation}

The goal of the agent is to learn a policy $\pi$ that maximises the value function.
\begin{defn}[Optimal policy] \label{def:optimal_policy}
    Let $\alpha \in (0,1]$ be the discount factor.
    The optimal policy $\pi^*$ is defined as the policy that maximises the value function:
    \begin{equation} \label{eq:optimal_policy}
        \pi^* = \argmax_{\pi} V_{\pi}(x)
    \end{equation}
\end{defn}

An important result in reinforcement learning is the Bellman equation \cite{bellman1952theory} which establishes the dynamic programming principle of optimality.
The Bellman equation states that the value function for the optimal policy can be expressed in terms of the immediate reward and the value function of the next state:
\begin{equation} \label{eq:bellman_equation}
    V_{\pi^*}(x) = \E \left[ R(x,a,x') + \alpha V_{\pi^*}(x') \right]
\end{equation}

It is often more convenient to work with the action-value function, also known as the $Q$-function, which in addition to the value function, assumes taking action $a$ in the starting state $x$ and then following the policy $\pi$:
\begin{equation}
        Q_\pi(x,a) := \E \left[ R(x,a,x') + \alpha V_\pi(x') \right]
\end{equation}
which together with Equation~\eqref{eq:bellman_equation} implies that we have
\begin{equation} \label{eq:q_v_equivalence}
    \max_{a} Q_{\pi^*}(x,a) = V_{\pi^*}(x)
\end{equation}
Together with the Bellman equation, we can define an optimal Bellman operator $\T$ for the $Q$-function:
\begin{equation} \label{eq:optimal_bellman_operator}
    \begin{split}
    \T Q_{\pi^*}(x,a) &:= \E \left[ R(x,a,x') + \alpha \max_{a'} Q_{\pi^*}(x',a') \right]
    \\
    &= Q_{\pi^*}(x,a)
    \end{split}
\end{equation}
Equation~\eqref{eq:optimal_bellman_operator} gives us a way to compute the optimal $Q$-function iteratively by finding the fixed point of the operator $\T$.
This is the basis of many reinforcement learning algorithms including the ones we will discuss in Section~\ref{sec:Algorithms}.

\subsubsection{Portfolio Optimisation Problem} \label{sec:PortfolioProblem}

We describe the key components of the portfolio optimisation problem in the context of the MDP framework.
First, the policy is paramaterised by a neural network $\pi_\theta$ with parameters $\theta$.
The policy can be stochastic where the neural network outputs parameters for a multivariate Gaussian distribution with a diagonal covariance matrix given the current state, $\pi_\theta: \X \rightarrow \mathbb{R}^{2n}$.
The action is then sampled from this distribution.

The state $x_t \in \X$ is a vector of:
\begin{enumerate}
  \item Historical prices of the stocks, $(S_{1,t-l+1}, \ldots, S_{n,t-l+1}, \ldots, S_{1,t}, \ldots, S_{n,t}) \in \R^{nl}$, of dimension $n \cdot l$ where $l$ is the number of time steps used in the historical window.
  \item Current portfolio weights before rebalancing $(w_1,\ldots,w_n) \in \R^n$ where $w_i$ is the percentage of wealth invested in the $i$-th stock.
  \item Current wealth, $W_t \in \R$.
\end{enumerate}

Therefore $x_t \in \R^m$ where $m=n(l + 1) + 1$.
We simulate prices for $l$ time steps before the start of the episode to provide the initial state.
If we use $w_0$ to denote the cash weight, then the sum of the weights must be equal to one: $\sum^n_{i=0} w_i = 1$.
Note that the weights are allowed to be negative to signify shorting of stocks or the borrowing of cash for a leveraged portfolio.
The action $a_t \in \A$ is a vector of the portfolio weights (excluding $w_0$), which is continuous, and is produced by the policy conditioned on the state.
Therefore, the weights at time step $t+1$ before rebalancing is exactly $a_t$.

The reward, $r_t:=R(x_t,a_t,x_{t+1})$ is the reward received after taking action $a_t$ in state $x_t$ and transitioning to state $x_{t+1}$.
We choose to use the Kelly criterion (log utility) as the objective function for the portfolio optimisation problem.
The Kelly criterion is extensively studied in \cite{maclean2011kelly} and is shown to have many desirable properties such as maximising the rate of capital growth and bankrupty avoidance (see \cite{ziemba2006good} for a summary).

With the Kelly criterion as the objective, the total accumulated reward, omitting the discount factor, is the log of the final wealth divided by initial wealth, $\sum^{N-1}_{t=0} r_t = \log{\frac{W(T)}{W(0)}}$.
Due to the additive properties of the log function, we can define the single period reward as the log of the wealth change at each time step as $r_t = \log{\frac{W_{t+1}}{W_t}}$.
In our current setting, the Kelly criterion also has the effect of equating our time averaged (undiscounted) objective with the time averaged growth rate of the wealth in the limit $T \to \infty$.
This means
\begin{equation}
    \frac{1}{T} \E \left[ \log{\frac{W(T)}{W(0)}} \right] = \lim_{T \to \infty} \frac{1}{T} \log{\frac{W(T)}{W(0)}}
\end{equation}
which can be seen immediately from Equation~\eqref{eq:wealth_sde} when we integrate the log wealth process over the time horizon $T$ and using the law of large numbers to give us $\frac{B(T)}{T} \to 0$ as $T \to \infty$ for any Brownian motion $B(t)$.
This is significant due to the following reason which is also discussed in \cite{peters2016evaluating,baumann2023reinforcement}.
An expected value is an ensemble average that is achieved when we are able to average over many independent realisations of the environment, which in physical terms means we reset the market to the same initial state and replay the episode.
This is clearly not possible in real financial markets as we only have one realisation of the market.
Therefore, what is more relevant is the time averaged growth rate of the wealth in a single realisation of the market.
The Kelly criterion allows to maximise the time averaged growth rate through the maximisation of the expected log wealth as the two converge in the long run.

The discount factor used in reinforcement learning has the effect of reducing the importance of future rewards.
This is analogous to the discounting of future cash flows in finance.
However, the use in reinforcement learning has a more practical purpose which helps to stabilise the learning process by reducing the variance of the gradient estimator or
  ensuring a finite sum of rewards in the case of infinite horizon problems \cite{sutton2018reinforcement}.
In this finite horizon setting, the discount factor can be set to 1 as the rewards are finite.
However, we exclude 0 as a discount factor as this would result in the agent only considering the immediate reward at each time step which turns the problem into single step optimisation.
When analysing the results, we will use the total accumulated reward without discounting divided by the time horizon which is analogous to a per annum exponential growth rate.

For the regime switching model, the problem becomes a partially observable Markov decision process (POMDP) \cite{littman2009tutorial} as the latent state $Z$ is not directly observable.
The variable $x_t$ is now an observable that is conditioned on the latent state $Z_t$.
The objective remains the same in this setting.

\section{Optimal Policy} \label{sec:OptimalPolicy}

In this section, we derive the optimal policy for the portfolio optimisation problem without market impact using the Kelly criterion as the objective.

\subsection{GBM without Market Impact} \label{sec:GBMbaseline}

The wealth at time $t$, $W(t)$ is the total value of the portfolio which consists of the stocks and cash.
We assume the cash account $V(t)$ accrues interest at a continuously compounded \emph{deterministic} interest rate $r_f$ which means it follows the SDE
\begin{equation}
    dV(t) = r_f V(t)dt
\end{equation}
Let $w_i$ be the weight of the $i$-th stock in the portfolio and $w_0 = 1 - \sum^n_{i=1} w_i$ be the weight of cash.
With zero market impact, the dynamics of the portfolio value is given by
\begin{equation}
    \begin{split}
        \frac{dW(t)}{W(t)} &= \left( 1-\sum^n_{i=1} w_i \right) \frac{dV(t)}{V(t)} + \sum^n_{i=1} w_i \frac{dS_i(t)}{S_i(t)} \\
        \\
    \end{split}
\end{equation}
After applying Ito's lemma, we can express the dynamics of the log wealth as
\begin{equation} \label{eq:wealth_sde}
    d\log W(t) = \left( 1-\sum^n_{i=1} w_i \right) r_f dt + \sum^n_{i=1} \left[ w_i \left( \mu_i dt + \sigma_i dB_i(t) \right) - \frac{1}{2} \sum^n_{j=1} w_i w_j \sigma_i \sigma_j \rho_{ij} dt \right].
\end{equation}
Noting that the $dB_i(t)$ terms are martingales, we can take the expectation of the log wealth process to get
\begin{equation}
    \E\left[d\log W(t)\right] = \left[ \left(1 - \sum^n_{i=1} w_i \right) r_f + \sum^n_{i=1} \left( w_i \mu_i - \frac{1}{2} \sum^n_{j=1} w_i w_j \sigma_i \sigma_j \rho_{ij} \right) \right] dt.
\end{equation}

Therefore, we have the following objective function:
\begin{equation}
    L(w) = \left(1 - \sum^n_{i=1} w_i \right) r_f + \sum^n_{i=1} \left[ w_i \mu_i - \frac{1}{2} \sum^n_{j=1} w_i w_j \sigma_i \sigma_j \rho_{ij} \right] \label{eq:objective}
\end{equation}

Maximising this objective results in a linear set of equations which is solved to obtain the optimal weights:
\begin{equation}
  \sum_{j=1}^{n}{\sigma_i \sigma_j \rho_{ij} w_j}=\mu_i-r \label{eq:optimal_weights}
\end{equation}
for $i=1,\ldots,n$.

Due to the assumption of a non-changing environment, the optimal policy is a \emph{fixed} weight portfolio, i.e., the weights do not change over time.
This means that the agent must learn to rebalance the portfolio back to the optimal weights after price movements have caused them to change.
However, if the environment is changing, the optimal weights will also change.
We explore this aspect in Section~\ref{sec:RSbaseline} with a regime switching model.

\subsubsection{Risk Preference}

\begin{figure}[h]
  \centering
  \includegraphics[scale=0.5]{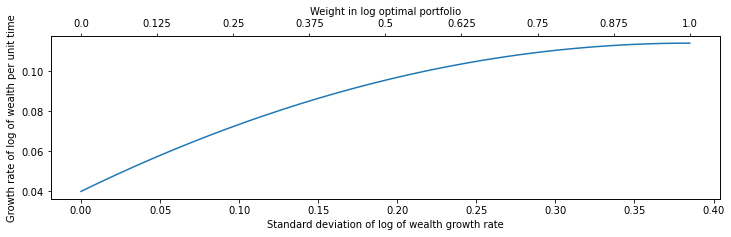}
  \caption{Efficient frontier for GBM without market impact.}
  \label{fig:eff_frontier}
\end{figure}

In terms of risk preference, we adopt the volatility of the change in log of wealth as a proxy measure of risk.
From Equation \eqref{eq:wealth_sde}, we can see that the volatility of the portfolio will be in the form of:
\begin{equation}
  \sum^n_{i=1} \sum^n_{j=1} w_i w_j \sigma_i \sigma_j \rho_{ij}
\end{equation}
Therefore, if we include a constraint on the volatility, the Lagrangian function has the same form as in Equation \eqref{eq:objective} except with a Lagrange multiplier for the last term.
In this case, the efficient frontier, under the assumption of no market impact,
  can be formed by a linear combination of the optimal portfolio derived from solving Equation \eqref{eq:optimal_weights} and cash.
In other words, we can focus on finding the log optimal portfolio and use it to obtain any portfolio on the efficient frontier by taking a fraction of the weights as shown in Figure~\ref{fig:eff_frontier}.
This is sometimes called a fractional Kelly strategy (see e.g.\cite{davis2011fractional}).

\subsection{GBM with Market Impact} \label{sec:MIbaseline}
The inclusion of market impact should cause the optimal policy to change.
We note that the optimal policy is the same when $\eta$ and $\gamma$, which control the market impact, are set to 0.
Therefore, when market impact is low, the optimal policy is close to the one without market impact.
If each period is sufficiently short, the change in prices inter-period would be small which results in small adjustments to the portfolio, and hence low market impact.
The exception is the start of the episode when the portfolio is all cash and large adjustments have to be made to bring the portfolio to the desired weights.
If the initial wealth is large, then this initial portfolio building phase can incur a large cost.
One way to mitigate this is to gradually adjust the portfolio weights to the required weights over the first $n$ periods.
If $n$ is small relative to the episode length, then the resulting accumulated reward would be close to the optimal level without market impact.

Therefore, we can use the same optimal policy as the one without market impact as a baseline when the initial wealth is low.
When the initial wealth is high, we slightly modify the baseline by staggering the initial portfolio build up.

\subsection{Regime Switching with Market Impact} \label{sec:RSbaseline}

With the regime switching model, the optimal weights in each regime can be very different which results in large adjustments to the portfolio.
Under low wealth scenarios, gradually adjusting the portfolio weights to the required weights over the first $n$ periods was still sufficient to get close to the optimal growth rate without market impact.
However, it was found to be insufficient under high wealth scenarios, producing extremely suboptimal growth rates.
In fact, a high rate of bankruptcies were observed without a sufficiently long adjustment period.
There are two reasons for this.
Firstly, the adjustments can now happen multiple times within an episode.
This causes costs to accumulate and the number of periods which the portfolio have suboptimal weights, due to the gradual adjustment, increases.
Secondly, the difference in weights between a bullish and bearish regime is far larger than the difference between an all cash portfolio and single regime optimal weights.

To mitigate this, we observe that the efficient frontier in each regime also consists of a linear combination of the optimal weights in that regime and cash.
A heuristically derived policy where a fraction of the optimal weights, in combination with gradual adjustments was used.
A grid search on the fraction and adjustment period was performed to find the baseline policy.
The performance of this policy is described in Section~\ref{sec:rs_highwealth}.

\section{Algorithms} \label{sec:Algorithms}

In this section, we will review five prominent model free DRL algorithms and the fundamental concepts behind them.
The algorithms are
\begin{enumerate}
  \item Advantage actor critic (A2C) (Variant of \cite{mnih2016asynchronous})
  \item Proximal policy optimisation (PPO) \cite{schulman2017proximal}
  \item Deep deterministic policy gradients (DDPG) \cite{lillicrap2016continuous}
  \item Twin delayed deep deterministic policy gradients (TD3) \cite{fujimoto2018addressing}
  \item Soft actor critic (SAC) \cite{haarnoja2018soft,haarnoja2018softb}
\end{enumerate}
The model free aspect means that the algorithms do not need to explicitly model the environment.
The algorithms learn a policy directly from the data collected from the environment.
All the algorithms are based on the actor-critic framework which we describe in Section~\ref{sec:actor_critic}.
The algorithms also all can operate in continuous action spaces which is better suited for the portfolio optimisation problem since the weights are continuous.
Some of the differences between the algorithms are in on-policy vs. off-policy learning which we describe in Section~\ref{sec:OnPolicyVsOffPolicy} and the type of policy representation (stochastic vs. deterministic) which we describe in Section~\ref{sec:policy_representation}.

The choice of the algorithms are motivated by their popularity in the literature and their ability to solve continuous control problems.
The implementations from the Stable Baselines 3 library \cite{raffin2021stable} were used for experiments.

\subsection{Actor-Critic Methods Overview} \label{sec:actor_critic}

Actor-Critic (AC) methods are a class of RL algorithms that maintain two distinct components, typically represented by parameterised function approximators (e.g., neural networks):
\begin{itemize}
    \item \textbf{Actor:} Represents the policy $\pi_\theta: \X \to \A$, responsible for selecting actions based on the current state. The parameters $\theta$ are updated to improve the policy's performance.
    \item \textbf{Critic:} Represents a value function, either $V_\phi: \X \to \R$ or $Q_\phi: \X \times \A \to \R$, responsible for evaluating the states or state-action pairs visited by the actor. The parameters $\phi$ are updated based on temporal difference (TD) learning principles \cite[Chapter 6]{sutton2018reinforcement}.
\end{itemize}
The core idea is to use the critic's evaluations to guide the updates of the actor's policy parameters.
The actor adjusts its action distribuion (or deterministic action) in directions that the critic suggests will lead to higher value.
This done through the policy gradient theorem, which provides a way to compute the gradient of the expected return with respect to the policy parameters.
There is the stochastic policy gradient theorem \cite{sutton1999policy} and the deterministic policy gradient theorem \cite{silver2014deterministic} which are used in the stochastic and deterministic actor-critic methods respectively.

This architecture aims to combine the strengths of policy-based methods (ability to handle continuous action spaces, direct policy optimisation) and value-based methods (improved sample efficiency through TD learning and potentially off-policy updates).

Actor-critic methods provide a flexible foundation upon which many advanced DRL algorithms are constructed.
The specific algorithms used in this paper (A2C, DDPG, PPO, SAC, TD3) are all instances of the actor-critic paradigm.
They differ primarily in their choices regarding:
\begin{itemize}
    \item Policy representation (stochastic vs. deterministic)
    \item Value function representation (V-function $V_\phi(x)$ vs. Q-function $Q_\phi(x,a)$)
    \item Method for value estimation (e.g., n-step returns, GAE, soft Bellman updates)
    \item Techniques used to stabilise the learning process and improve efficiency (e.g., target networks, replay buffers, policy constraints, entropy maximisation)
\end{itemize}
These variations represent different strategies tailored to address specific challenges encountered in DRL, such as learning in continuous action spaces, ensuring stable convergence, promoting efficient exploration, and maximising sample efficiency.

\subsection{On-Policy vs. Off-Policy} \label{sec:OnPolicyVsOffPolicy}
\begin{itemize}
    \item \textbf{On-Policy (A2C, PPO):} These algorithms require that the data used for policy updates is collected using the exact same policy that is being optimised. This typically means that after each update phase, the collected data is discarded, and new data must be generated with the updated policy.
    \item \textbf{Off-Policy (DDPG, SAC, TD3):} These algorithms can learn a policy using data generated by a different policy (sometimes called the behaviour policy), typically older versions of the current policy stored in a replay buffer.
\end{itemize}
On-policy approaches often leads to more stable convergence, as the updates are directly related to the performance of the current policy. However, it is inherently sample inefficient because data cannot be reused extensively.
Exploration is also directly tied to the current policy's behavior (e.g., through its stochasticity).
Off-policy learning enables significantly better sample efficiency through the reuse of past experiences stored in the replay buffer.
This is particularly advantageous in scenarios where environment interaction is costly (e.g., trading in the real financial markets).
However, learning from off-policy data can introduce instability due to the mismatch between the data distribution under the behavior policy and that of the current policy.
This necessitates mechanisms like importance sampling (implicitly handled in $Q$-learning updates) or careful stabilisation techniques (target networks, conservative updates).

The choice between on-policy and off-policy often hinges on the cost of data collection versus the need for stability.
Off-policy methods are generally favored when interactions are expensive, while on-policy methods are preferred when interactions are cheap or stability is paramount.

\subsection{Policy Representation} \label{sec:policy_representation}
\begin{itemize}
    \item \textbf{Deterministic (DDPG, TD3):} The actor network (policy) outputs a single, specific action for a given state.
    \item \textbf{Stochastic (A2C, PPO, SAC):} The actor network outputs a probability distribution over actions. Actions are then are sampled from this distribution during interaction.
\end{itemize}

Deterministic policies simplify the critic update (no expectation over actions needed in the target value calculation).
However, exploration must be introduced externally, typically by adding noise to the actor's output during training.
This separation can sometimes make exploration less systematic.
Exploration is inherently part of a stochastic policy's behavior through sampling.
Stochastic policies can naturally represent uncertainty and potentially learn multi-modal behaviors (i.e., multiple good actions in a given state) and offer a more integrated approach to exploration.

\subsection{Advantage Actor-Critic (A2C)}
Advantage Actor-Critic (A2C) is fundamentally a synchronous, deterministic implementation variant of the Asynchronous Advantage Actor Critic (A3C) algorithm introduced in \cite{mnih2016asynchronous}.
A3C was a pioneering DRL algorithm that utilised multiple parallel actors, each interacting with its own copy of the environment and asynchronously calculating gradient updates for the neural network parameters.
This asynchronous nature was intended to decorrelate the agents' data and stabilise learning.

However, asynchronous updates can introduce challenges.
Workers, which are independent copies of the environment and agent, compute gradients based on slightly outdated versions of the policy, which leads to some noise in the updates and it is unclear whether this noise is beneficial or detrimental to learning.
Furthermore, the asynchronous communication pattern might not fully leverage the parallel processing capabilities of modern hardware like GPUs, which often perform best with large, batched computations (see e.g. \cite{wu2017openai}).

A2C addresses these points by synchronising the update step.
Multiple actors still run in parallel, collecting experience from their respective environment instances.
However, instead of updating the global network asynchronously, A2C waits for all workers to complete a predefined number of steps (or episodes) before aggregating their experiences.
Gradients are computed based on this larger batch of data, and a single, synchronous update is applied to the neural network parameters.

Empirical results, particularly from OpenAI Baselines, indicated that A2C often achieved comparable or even superior performance to A3C, suggesting that the benefits of synchronous updates outweighed the potential exploration advantages of asynchrony in many common benchmark tasks (see \cite{wu2017openai}).

\subsubsection{Policy and Value Representation}
\begin{itemize}
    \item \textbf{Actor (Policy) network  $\pi_\theta$:} The actor network outputs a stochastic policy. For discrete action spaces, it typically outputs logits which are passed through a softmax function to produce a probability distribution over actions. For continuous action spaces, it usually outputs the parameters of a probability distribution, most commonly the mean and standard deviation of a Gaussian distribution. Actions are then sampled from this distribution during interaction.
    \item \textbf{Critic (Value) network $V_\phi$:} The critic network approximates the value function and learns to evaluate the expected return from a state $x$ under the current policy.
\end{itemize}

\subsubsection{Advantage Estimation}
A core component of A2C is the use of the advantage function $A_\pi(x, a) = Q_\pi(x, a) - V_\pi(x)$ where the superscripts denote the policy under which the value function is computed.
The advantage quantifies how much better or worse a specific action $a$ is compared to the average action taken from state $s$ under policy $\pi$.
Using the advantage estimate $\hat{A}_t$ instead of the raw rewards in the policy gradient calculation significantly reduces the variance of the gradient estimate, leading to more stable learning.
The state value $V_\pi(x)$ acts as a baseline.

In practice, $Q_\pi(x, a)$ is not directly estimated.
Instead, A2C typically uses $n$-step rewards as a target for the value function and implicitly for the advantage calculation.
The $n$-step rewards provides a target for $V_\phi(x_t)$:
\begin{equation}
    V_{\text{target}}(x_t) := \sum_{s=t}^{t+n-1} \alpha^{s-t} R(x_s,a_s,x_{s+1}) + \alpha^{n} V_\phi(x_{t+n})
\end{equation}
where $V_\phi(x_{t+n})$ is the critic's estimate of the value of the state reached after $n$ steps.
The advantage estimate $\hat{A}_t$ for the state-action pair $(x_t, a_t)$ is then calculated as the difference between this $n$-step target and the critic's estimate of the value of the current state:
\begin{equation}
    \hat{A}_t = V_{\text{target}}(x_t) - V_\phi(x_t)
\end{equation}

\subsubsection{Loss Functions and Optimisation}
The network parameters $(\theta, \phi)$ are updated by minimising a combined loss function, typically composed of three terms:
\begin{enumerate}
    \item \textbf{Critic Loss (Value Loss):} This loss drives the critic $V_\phi$ to accurately predict the $n$-step target returns. It is usually the Mean Squared Error (MSE) between the predicted values and the calculated target values
    \begin{equation} \label{eq:a2c_critic_loss}
        L_V(\phi) = \frac{1}{N_B} \sum_{i=1}^{N_B} \left[ \left( V_{\text{target}}(x_t^{(i)}) - V_\phi(x_t^{(i)}) \right)^2 \right]
    \end{equation}
    where $N_B$ is the batch size and $x_t^{(i)}$ is the state at step $t$ in the $i$-th sample in the batch.

    \item \textbf{Actor Loss (Policy Loss):} This is the standard policy gradient objective, aiming to increase the log-probability of actions that led to higher-than-expected returns (positive advantage) and decrease the log-probability of actions that led to lower-than-expected returns (negative advantage).
    The advantage estimate $\hat{A}_t$ is used to scale the gradient:
    \begin{equation} \label{eq:a2c_actor_loss}
        L_\pi(\theta) = -\frac{1}{N_B} \sum_{i=1}^{N_B} \log \pi_\theta(a_t^{(i)} | x_t^{(i)}) \hat{A}_t^{(i)}
    \end{equation}
    where $\pi_\theta(a_t^{(i)} | x_t^{(i)})$ is the probability of taking action $a_t^{(i)}$ in state $x_t^{(i)}$ under the current policy $\pi_\theta$.
    Crucially, the advantage estimate $\hat{A}_t$ is treated as a fixed baseline when computing the gradient with respect to the actor parameters $\theta$. The negative sign indicates that we perform gradient ascent on the expected return objective.

    \item \textbf{Entropy Loss:} An optional but commonly used term that encourages exploration by penalising policies with low entropy (i.e., policies that are too deterministic).
    Maximising entropy encourages the policy to assign non-zero probabilities to more actions, preventing premature convergence to suboptimal deterministic policies.
    The entropy loss is defined as:
    \begin{equation} \label{eq:a2c_entropy_loss}
        L_H(\theta) = -\sum_{i=1}^{N_B} H(\pi_\theta(\cdot | x_t^{(i)}))
    \end{equation}
    where $H$ denotes the entropy of the policy distribution at state $x_t^{(i)}$ and is computed depending on the type of distribution used (e.g., Gaussian).
\end{enumerate}
The total loss function is a weighted sum of these components:
\begin{equation} \label{eq:a2c_total_loss}
    L(\theta, \phi) = L_\pi(\theta) + c_1 L_V(\phi) + c_2 L_H(\theta)
\end{equation}
where $c_1$ and $c_2$ are coefficients controlling the relative importance of the value loss and the entropy loss.

\subsection{Proximal Policy Optimisation (PPO)}
Proximal Policy Optimisation (PPO) \cite{schulman2017proximal} is an on-policy actor-critic algorithm using heuristics inspired by the ideas from Trust Region Policy Optimisation (TRPO) \cite{schulman2015trust}.
Standard policy gradient algorithms can be sensitive to step size; a single large update based on a batch of data can drastically change the policy, potentially leading to a collapse in performance.

PPO addresses this by aiming to take the largest possible improvement step on the policy at each iteration, without moving too far from the previous policy, thereby reducing the risk of performance collapse.
It achieves this by optimising a surrogate objective function that incorporates a mechanism to penalise large changes in the policy distribution.
This philosophy is similar to TRPO, which enforces a hard constraint on the KL divergence between the old and new policies.
However, TRPO involves complex second-order optimisation methods (like the conjugate gradient algorithm) to solve the constrained optimisation problem.
PPO was developed as a simpler alternative that achieves similar stability benefits using only first-order optimisation, making it much easier to implement, parallelise, and apply to large-scale problems.
The actor and critic networks in PPO are similar to those in A2C, but the training process and objective functions differ in a meaningful way.

\subsubsection{Surrogate Objective Functions} \label{sec:ppo_surrogate_objective}
The core innovation of PPO lies in its surrogate objective functions, which modify the standard policy gradient objective as we saw in \eqref{eq:a2c_actor_loss}.
Taking $\pi_\theta(a_t | x_t)$ to be the probability of taking action $a_t$ in state $x_t$ under the current policy $\pi_\theta$, PPO aims to optimise an objective related to
\begin{equation}
    \begin{split}
    L^{CPI}(\theta) &= \E[ \frac{\pi_\theta(a_t | x_t)}{\pi_{\theta_{old}}(a_t | x_t)} \hat{A}_t ]
    \\
    &= \E \left[ k_t(\theta) \hat{A}_t \right]
    \end{split}
\end{equation}
where $\pi_{\theta_{old}}$ is the policy used to collect the data (the "old" policy) and $k_t(\theta) = \frac{\pi_\theta(a_t | x_t)}{\pi_{\theta_{old}}(a_t | x_t)}$ is the probability ratio between the new and old policies for the action taken at time $t$.
Optimising $L^{CPI}$ directly can lead to excessively large updates when $k_t(\theta)$ deviates significantly from 1.
PPO introduces modifications to discourage large ratios.

There are two main variants of the PPO surrogate objective.
We focus on the more commonly used clipped version, which is simpler and has shown strong empirical performance across a wide range of tasks \cite{andrychowicz2021what,engstrom2020implementation}.
The objective is defined as
\begin{equation} \label{eq:ppo_clip_objective}
    L^{CLIP}(\theta) = \E \left[ \min \left( k_t(\theta) \hat{A}_t, \text{clip}(k_t(\theta), 1 - \epsilon, 1 + \epsilon) \hat{A}_t \right) \right]
\end{equation}
Here, $\epsilon$ is a small hyperparameter that defines the clipping range.
The $\text{clip}(k_t(\theta), 1 - \epsilon, 1 + \epsilon)$ function constrains the probability ratio $k_t(\theta)$ to lie within the interval $[1 - \epsilon, 1 + \epsilon]$.
The $\min$ operator ensures that the final objective is a lower bound (a pessimistic estimate) of the unclipped objective $k_t(\theta) \hat{A}_t$.
\begin{itemize}
    \item If the advantage $\hat{A}_t$ is positive (action was better than average), the objective increases as $k_t(\theta)$ increases (making the action more likely), but the increase is capped once $r_t(\theta)$ exceeds $1 + \epsilon$.
    \item If the advantage $\hat{A}_t$ is negative (action was worse than average), the objective increases as $k_t(\theta)$ decreases (making the action less likely), but the decrease is capped once $k_t(\theta)$ goes below $1 - \epsilon$.
\end{itemize}
Overall, this prevents the policy from moving too far from the old policy in a single update based on a potentially large positive or negative advantage estimate.

The clipping mechanism in PPO-Clip provides a remarkably simple yet effective way to implement the trust region concept using only first-order gradients.
It avoids the need for calculating Fisher-vector products or solving linear systems via conjugate gradient, which are required in TRPO.
This simplicity is a major reason for PPO's widespread adoption.

\subsubsection{Generalised Advantage Estimation} \label{sec:ppo_gae}
PPO typically employs Generalised Advantage Estimation (GAE) \cite{schulman2016high} to compute the advantage estimates $\hat{A}_t$ used in the surrogate objective.
GAE provides a sophisticated way to balance the trade-off between bias and variance in advantage estimation.
If we define an $n$-step advantage estimate $\hat{A}_t^{(n)}$ as
\begin{equation}
    \hat{A}_t^{(n)} := r_t + \alpha r_{t+1} + \alpha^2 r_{t+2} + \ldots + \alpha^{n-1} r_{t+n-1} + \alpha^n V(x_{t+n}) - V(x_t)
\end{equation}
Then GAE can be expressed as an exponentially weighted sum of the $n$-step advantage estimates
\begin{equation} \label{eq:gae}
    \begin{split}
    \hat{A}_t^{GAE}(\alpha, \lambda) &= (1 - \lambda)\left(\hat{A}_t^{(1)} + \lambda \hat{A}_t^{(2)} + \lambda^2 \hat{A}_t^{(3)} + \ldots\right)
    \\
    &= (1 - \lambda)\left(\delta_t + \lambda \left( \delta_t + \alpha \delta_{t+1} \right) + \lambda^2 \left( \delta_t + \alpha \delta_{t+1} + \alpha^2 \delta_{t+2} \right) + \ldots\right)
    \\
    &= (1 - \lambda)\left(\delta_t \left(1 + \lambda + \lambda^2 + \ldots\right) + \alpha \delta_{t+1} \left(\lambda + \lambda^2 + \ldots\right) + \alpha^2 \delta_{t+2} \left(\lambda^2 + \ldots\right) + \ldots\right)
    \\
    &= (1 - \lambda)\left(\delta_t \frac{1}{1 - \lambda} + \alpha \delta_{t+1} \frac{\lambda}{1 - \lambda} + \alpha^2 \delta_{t+2} \frac{\lambda^2}{1 - \lambda} + \ldots\right)
    \\
    &= \sum_{s=0}^{\infty} \left(\alpha \lambda\right)^s \delta_{t+s}
    \end{split}
\end{equation}
where $\lambda$ controls the exponential weighting and $\delta_t = r_t + \alpha V(x_{t+1}) - V(x_t)$ is the one-step temporal difference (TD) error.
We can see that with small values of $n$ in $\hat{A}_t^{(n)}$, the advantage estimate has low variance but high bias, as it relies more on the critic's estimate of the value function $V(x_t)$.
Conversely, with larger values of $n$, the advantage estimate has low bias but high variance, as it relies more on the empirical returns from the environment.
Therefore, the parameter $\lambda$ controls the balance between bias and variance in the GAE estimate by modulating the contribution of the $n$-step advantage estimates.
At the extremes, we have
\begin{itemize}
    \item $\lambda = 0$: This reduces to the one-step TD error, $\hat{A}_t = \delta_t = r_t + \alpha V(x_{t+1}) - V(x_t)$, which has low variance but potentially high bias if the value function estimate $V_\phi$ is inaccurate.
    \item $\lambda = 1$: We have telescoping sums when we expand the $n$-step advantage estimates, leading to the Monte Carlo advantage estimate, $\hat{A}_t = \sum_{l=0}^{\infty} (\alpha)^l r_{t+l} - V(x_t)$, which has low bias but can have very high variance, especially for long episodes.
\end{itemize}
Intermediate values of $\lambda$ (e.g., $\lambda = 0.95$) often provide a good compromise, reducing variance compared to Monte Carlo estimates while incurring less bias than one-step TD estimates, leading to more stable and efficient learning.
For finite time horizons, GAE is computed by truncating the final sum in \eqref{eq:gae}.

\subsubsection{Loss Functions and Optimisation}
The overall optimisation process in PPO involves minimising a combined loss function, typically including the policy surrogate objective, a value function loss, and an optional entropy bonus.
\begin{enumerate}
    \item \textbf{Value Loss:} The value function can be trained in a similar manner to A2C, using the mean squared error (MSE) between the predicted value and the target value as in \eqref{eq:a2c_critic_loss}.
    \item \textbf{Policy Loss:} This is the negative of the PPO surrogate objective $L^{CLIP}$ in \eqref{eq:ppo_clip_objective}
    \begin{equation} \label{eq:ppo_policy_loss}
        L_\pi(\theta) = - L^{CLIP}(\theta)
    \end{equation}
    \item \textbf{Entropy Loss:} Also similar to A2C, an entropy term can subtracted from the objective to encourage exploration as in \eqref{eq:a2c_entropy_loss}.
\end{enumerate}
The final loss function is a weighted sum of these components as in \eqref{eq:a2c_total_loss}.
A key feature of PPO is that it can perform multiple epochs of gradient updates on the same batch of collected data due to the clipped objective limiting the potential of large deviations from the old policy.
This improves data utilisation compared to standard on-policy gradient methods that perform only one update per sample.

\subsection{Deep Deterministic Policy Gradient (DDPG)}
Deep Deterministic Policy Gradient (DDPG) \cite{lillicrap2016continuous} is an off-policy algorithm specifically designed to operate in environments with continuous action spaces.
It adapts the ideas underlying the success of Deep Q-Networks (DQN), introduced in \cite{mnih2015human}, for discrete action spaces to the continuous domain.

\subsubsection{Policy and Value Representation}
\begin{itemize}
    \item \textbf{Actor Network $\pi_\theta$:} This network takes the current state $x$ as input and outputs a deterministic action $a = \pi_\theta(x)$.
    \item \textbf{Critic Network $Q_\phi$:} This network approximates the $Q$ function and learns to evaluate the expected return of taking action $a$ in state $x$ and following the actor's policy thereafter.
\end{itemize}

Standard $Q$-learning \cite{watkins1992q} involves a maximisation step over actions $\max_{a} Q(x,a)$ when computing the target value, which is straightforward in discrete spaces but intractable in continuous spaces when the $Q$-function is approximated by a neural network with a large number of parameters.

DDPG circumvents this issue by using the actor network in place of the maximisation step.
This deterministic actor provides a specific action $a_t$ for the next state $x_{t+1}$, allowing the critic $Q_\phi(x_t, a_t)$ to be updated using a target value $Q_{\text{target}} = R(x_t,a_t,x_{t+1}) + \alpha Q_\phi(x_{t+1}, \pi_\theta(x_{t+1}))$ without needing an explicit maximisation over the continuous action space.
The algorithm builds upon the Deterministic Policy Gradient theorem from \cite{silver2014deterministic}, which shows how to compute the policy gradient for deterministic policies using the critic network $Q_\phi$.
Specifically, for the objective
\begin{equation} \label{eq:policy_objective}
    J(\theta) = \E \left[ Q_\phi(x, \pi_\theta(x)) \right]
\end{equation}
we can compute the gradient with respect to the policy parameters $\phi$ as
\begin{equation} \label{eq:deterministic_policy_gradient}
    \nabla_\theta J(\theta) = \E \left[ \nabla_\theta Q_\phi(x, \pi_\theta(x)) \right]
\end{equation}
This allows the actor to be updated in the direction that increases the expected return, as estimated by the critic, hence it is a substitute for the maximisation step in standard $Q$-learning.
Additionally, DDPG utilises target networks for both the actor and the critic to stabilise learning, as detailed in the next section.

\subsubsection{Off-Policy Learning: Replay Buffer and Target Networks}
DDPG leverages two key components crucial for off-policy learning and stability:
\begin{itemize}
    \item \textbf{Replay Buffer:} Like DQN, DDPG uses a replay buffer (experience replay memory) to store past transitions $(x_t, a_t, r_t, x_{t+1})$ collected during interaction with the environment.
    During training updates, mini-batches of transitions are randomly sampled from this buffer.
    This technique breaks the temporal correlations inherent in sequential experience, leading to less bias in the gradient computations (see \cite[Chapter 8.1.3]{goodfellow2016deep}).
    The reuse of data in the replay buffer also leads to higher data efficiency.

    \item \textbf{Target Networks:} DDPG maintains separate target networks, $Q_{\phi'}$ and $\pi_{\theta'}$, which are copies of the main critic and actor networks, respectively.
    These target networks are used exclusively for calculating the target values needed for the critic update.
    The parameters of the target networks $(\theta', \phi')$ are not updated by backpropagation directly.
    Instead, they are updated slowly to track the main network weights $(\theta, \phi)$ either by hard updates where the parameters are copied directly after a certain number of updates to the main parameters or by soft updates where the target network parameters are updated using a small fraction of the main network parameters:
    \begin{equation} \label{eq:ddpg_target_network_update}
        \begin{split}
            \theta' &\leftarrow \tau \theta + (1 - \tau) \theta'
            \\
            \phi' &\leftarrow \tau \phi + (1 - \tau) \phi'
        \end{split}
    \end{equation}
    where $\tau$ is a small constant (e.g., 0.001).
\end{itemize}
The use of target networks is critical for stabilising the learning process in $Q$-learning-based methods like DDPG.
The critic update aims to minimise the difference between $Q_\phi(x, a)$ and a target value $Q_{\text{target}} = r_t + \alpha Q_{\phi'}(x_{t+1}, \pi_{\theta'}(x_{t+1}))$.
If the main networks $Q_\phi$ and $\pi_\theta$ were used directly to compute $Q_{\text{target}}$, the target value itself would depend on the parameters $\theta$ and $\phi$ that are being actively optimised.
This creates a "moving target" problem, where the optimisation objective changes rapidly at each step, often leading to oscillations or divergence.
By using slowly updated target networks $Q_{\phi'}$ and $\pi_{\theta'}$, the target value $Q_{\text{target}}$ becomes more stable over short timescales, providing a consistent objective for the critic update and preventing runaway feedback loops.

\subsubsection{Loss Functions and Optimisation}
Updates are performed by sampling a mini-batch of transitions $(x_t, a_t, r_t, x_{t+1})$ from the replay buffer.
\begin{enumerate}
    \item \textbf{Critic Loss:} The critic network $Q_\phi(x, a)$ is updated by minimising the Mean Squared Error (MSE) loss between its output and the target value $Q_{\text{target}}$:
    \begin{equation}
        L_Q(\phi) = \frac{1}{N_B} \sum_{i=1}^{N_B} \left( Q_{\text{target}}^{(i)} - Q_\phi(x_t^{(i)}, a_t^{(i)}) \right)^2
    \end{equation}
    where $N_B$ is the batch size and $Q_{\text{target}}^{(i)} = r_t^{(i)} + \alpha Q_{\phi'}(x_{t+1}^{(i)}, \pi_{\theta'}(x_{t+1}^{(i)}))$.
    The critic parameters $\phi$ are updated using gradient descent on this loss.

    \item \textbf{Actor Loss:} The actor network $\pi_\theta(x)$ is updated by maximising the expected return as estimated by the critic $Q_\phi(x, a)$.
    This is done by ascending the policy gradient derived from the deterministic policy gradient theorem:
    \begin{equation} \label{eq:ddpg_actor_loss}
        L_\pi(\theta) = -\frac{1}{N_B} \sum_{i=1}^{N_B} Q_\phi(x_t^{(i)}, \pi_\theta(x_t^{(i)}))
    \end{equation}
    The negative sign indicates that we perform gradient ascent on the expected return objective.
    Note that the critic parameters $\phi$ are held constant during the actor update calculation.

    \item \textbf{Target Network Updates:} After the actor and critic updates, the target network weights are updated using the soft update rule from \eqref{eq:ddpg_target_network_update}.
\end{enumerate}

Since the policy $\pi_\theta(x)$ is deterministic, exploration needs to be added explicitly during training.
This is typically done by adding noise to the actor's output action before executing it in the environment.
A common choice is temporally correlated noise, such as an Ornstein-Uhlenbeck process, or simpler uncorrelated Gaussian noise.
The scale of the noise is often annealed over the course of training to reduce exploration as the policy becomes more refined.

Despite its theoretical appeal for continuous control, DDPG is notoriously sensitive to hyperparameters and can suffer from instability \cite{duan2016benchmarking,henderson2019deep,haarnoja2018soft}.
A primarily reason for this fragility is the tendency of the critic to overestimate $Q$-values, particularly due to function approximation errors \cite{thrun1993issues,hasselt2010double}.
The deterministic actor is trained to output actions that maximise the (potentially overestimated) $Q$-values from the critic.
If the critic erroneously assigns a high value to a certain region of the state-action space, the actor will quickly learn to exploit this peak.
The critic's target calculation, $Q_{\text{target}} = r_t + \alpha Q_{\phi'}(x_{t+1}, \pi_{\theta'}(x_{t+1}))$, then uses the actor's action $\pi_{\theta'}(x_{t+1})$, potentially reinforcing the overestimation if $\pi_{\theta'}$ also targets the overestimated region.
This positive feedback loop between the actor and critic can lead to divergence or convergence to poor policies, motivating the improvements introduced in TD3.

\subsection{Twin Delayed Deep Deterministic Policy Gradient (TD3)}
Twin Delayed Deep Deterministic Policy Gradient (TD3) \cite{fujimoto2018addressing} directly addresses the primary shortcomings observed in DDPG, namely the overestimation of $Q$-values due to function approximation errors and the resulting high variance and instability in learning.
The core problem identified is that function approximation errors in the critic's $Q$-value estimates can lead to systematic overestimation, particularly when combined with the maximisation step implicit in the actor's update and the target value calculation.
TD3 introduces three specific modifications to the DDPG framework designed to mitigate these issues and improve performance and stability in continuous control tasks.
\begin{itemize}
    \item Clipped Double $Q$-Learning
    \item Delayed Policy Updates
    \item Target Policy Smoothing
\end{itemize}

\subsubsection{Clipped Double \texorpdfstring{$Q$}{}-learning}
To combat $Q$-value overestimation, TD3 employs a technique inspired by Double $Q$-learning.
It learns two independent $Q$-function networks (critics), $Q_{\phi_1}$ and $Q_{\phi_2}$, alongside their corresponding target networks, $Q_{\phi'_1}$ and $Q_{\phi'_2}$.
There is a single actor network $\pi_\theta$ with a target actor network $\pi_{\theta'}$ that is shared between the two critics and is updated using the gradients from one of the critics (typically the first one, $Q_{\phi_1}$).
When calculating the target value $Q_{\text{target}}$ for the update of both critics, TD3 uses the minimum of the values predicted by the two target $Q$-networks:
\begin{equation} \label{eq:td3_target}
    Q_{\text{target}} = r_t + \alpha \min_{i=1,2} Q_{\phi'_i}(x_{t+1}, \pi_{\theta'}(x_{t+1}))
\end{equation}

Using the minimum of the two $Q$-estimates provides a more conservative target value.
Since function approximation errors might lead one critic to overestimate the value for a given $(x,a)$, taking the minimum makes it less likely that this positive bias propagates through the Bellman updates.
The double $Q$-learning version of DQN \cite{vanhasselt2016deep} alternates between two $Q$ networks for evaluating and selecting the action, i.e., one network is used to calculate $Q$-values while the other is used to select the best action.
However, TD3 uses the minimum  with slowly changing policies, to prevent the situation where both critics might eventually learn to overestimate the values for actions chosen by the actor.
The minimum provides a consistent mechanism to push value estimates downwards, counteracting the overestimation bias inherent in $Q$-learning with function approximation.

\subsubsection{Delayed Policy Updates}
TD3 introduces a delay in updating the actor network $\pi_\theta$ and the target networks ($\pi_{\theta'}, Q_{\phi'_1}, Q_{\phi'_2}$) relative to the critic networks ($Q_{\phi_1}, Q_{\phi_2}$).
Typically, the critics are updated at every training step, while the actor and target networks are updated only once for every $d>1$ critic updates.

The rationale behind this delay is to reduce variance and improve stability.
The actor update relies on the $Q$-function estimates to compute the policy gradient, which is used to update the actor's parameters.
If the actor is updated too frequently, it can lead to high variance in the policy gradient estimates, especially if the $Q$-function estimates are still noisy or not well-converged.
By updating the critics more frequently than the actor, TD3 allows the $Q$-value estimates to stabilise and converge towards a more accurate representation of the current policy's value before that information is used to update the policy itself.
This decoupling leads to higher quality policy updates based on more reliable value estimates, reducing overall variance and stabilising the learning process.

\subsubsection{Target Policy Smoothing}
To further reduce variance and prevent the actor from exploiting sharp peaks in the $Q$-function estimate, TD3 introduces target policy smoothing regularisation.
When computing the action $a_{t+1}$ used in the target calculation $Q_{\text{target}} = r_t + \alpha \min_{i=1,2} Q_{\phi'_i}(x_{t+1}, a_{t+1})$, TD3 adds clipped random noise to the action selected by the target actor $\pi_{\theta'}$ with
\begin{equation}
    a_{t+1} = \left( \pi_{\theta'}(x_{t+1}) + \varepsilon \right)
\end{equation}
where $\varepsilon \sim \text{clip}(\mathcal{N}(0, \sigma), -c, c)$ for some $\sigma, c > 0$.
In other words, some Gaussian noise $\varepsilon$ is added to the action output by the target actor, but this noise is clipped to be within a range $[-c, c]$.

This technique acts as a regulariser by smoothing the value estimate along the action dimension.
Instead of bootstrapping off the value of a single point action $\pi_{\theta'}(x)$, the target value $Q_{\text{target}}$
effectively averages the minimum $Q$-value over a small neighborhood of actions around $\pi_{\theta'}(x)$.
This encourages the learned $Q$-functions ($Q_{\phi_1}, Q_{\phi_2}$) to be smoother with respect to actions, making them less prone to developing sharp, narrow peaks that might arise purely from function approximation errors.
A smoother value landscape makes the policy itself more robust, as small errors in the actor's output are less likely to lead to large changes in the estimated value.

\subsubsection{Loss Functions and Optimisation}
The TD3 update procedure, using mini-batches sampled from a replay buffer, proceeds as follows:
\begin{enumerate}
    \item \textbf{Compute Target $Q$-value with Clipped Double $Q$-learning and Target Action Smoothing:}
    Calculate $Q_{\text{target}}$ using the minimum of the two target critics and the smoothed target action
    \begin{equation}
        Q_{\text{target}} = r_t + \alpha \min_{i=1,2} Q_{\phi'_i}(x_{t+1}, \pi_{\theta'}(x_{t+1}) + \varepsilon)
    \end{equation}
    where $\varepsilon \sim \text{clip}(\mathcal{N}(0, \sigma), -c, c)$.
    \item \textbf{Critic Loss:} Update both critic networks $Q_{\phi_1}$ and $Q_{\phi_2}$ by minimising the MSE loss with respect to the target:
    \begin{equation}
        L_Q(\phi_j) = \frac{1}{N_B} \sum_{i=1}^{N_B} \left( Q_{\text{target}}^{(i)} - Q_{\phi_j}(x_t^{(i)}, a_t^{(i)}) \right)^2 \quad \text{for } j = 1, 2
    \end{equation}
    where $N_B$ is the batch size.
    \item \textbf{Delayed Actor and Target Updates:} Every $d$ critic updates, update the actor and target networks in the same manner as DDPG in \eqref{eq:ddpg_actor_loss} and \eqref{eq:ddpg_target_network_update} respectively.
\end{enumerate}

Similar to DDPG, TD3 requires exploration by adding noise to the action during data collection.
While there are more hyperparameters to tune in TD3 compared to DDPG, the algorithm is generally more robust and stable.

\subsection{Soft Actor-Critic (SAC)}
Soft Actor-Critic (SAC) \cite{haarnoja2018soft,haarnoja2018softb} is an off-policy actor-critic algorithm grounded in the maximum entropy reinforcement learning framework.
Unlike traditional RL algorithms that solely aim to maximise the expected cumulative reward, SAC modifies the objective to maximise a weighted sum of the expected return and the entropy of the policy at \emph{each} visited state.
The objective is modified to
\begin{equation} \label{eq:sac_objective}
    J(\pi) = \E \left[ \sum_{t=0}^{N} r(x_t, a_t) + \gamma H(\pi(\cdot | x_t)) \right]
\end{equation}
where $H(\pi(\cdot | x_t)) = \E [-\log \pi(a | x_t)]$ is the entropy of the policy $\pi$ at state $x_t$, and $\gamma > 0$ is the temperature parameter that determines the relative importance of the entropy term versus the reward.

The core motivation behind this objective is twofold:
\begin{itemize}
    \item \textbf{Enhanced Exploration:} By explicitly rewarding higher entropy, the policy is encouraged to explore more broadly, assigning non-negligible probabilities to multiple viable actions rather than collapsing prematurely to a single deterministic strategy. This can lead to the discovery of better solutions, especially in complex environments with multimodal reward landscapes.
    \item \textbf{Improved Robustness:} Policies learned under the maximum entropy objective tend to be more robust. They are incentivized to succeed while acting as randomly as possible, which can lead to smoother control and better generalisation to slight variations in the environment or task.
\end{itemize}

SAC combines this maximum entropy framework with an efficient off-policy actor-critic structure, aiming to achieve both high sample efficiency (through data reuse via a replay buffer) and stable, robust learning.
The maximum entropy objective fundamentally alters the RL goal from finding emph{the} optimal action to finding a policy that assigns high probability to \emph{all} good actions.
This intrinsic drive for exploration and stochasticity contrasts with methods like DDPG or TD3 that rely on externally added noise to a deterministic policy, or on-policy methods where exploration is coupled with the policy being optimised.

\subsubsection{Policy and Value Representation}
SAC employs several network components:
\begin{itemize}
    \item \textbf{Actor network $\pi_\theta$:} The actor network outputs the parameters of a stochastic policy distribution similar to A2C and PPO.
    \item \textbf{Critic Networks $Q_{\phi_1}, Q_{\phi_2}$:} SAC utilises two independent critic networks that estimate the \emph{soft} action-value function, which incorporates the entropy bonus.
    Learning two Q-functions helps mitigate the overestimation bias that can arise from function approximation errors similar to the Double $Q$-learning technique used in TD3.
    \item \textbf{Target Critic Networks $Q_{\phi'_1}, Q_{\phi'_2}$:} Corresponding target networks are maintained for each Q-network.
    These are updated slowly using soft updates (like in DDPG or TD3) and provide stable targets for the Q-function updates.
\end{itemize}

\subsubsection{Entropy-Augmented Objectives}
The maximum entropy objective leads to modified Bellman equations for the value functions.
The soft action-value function $Q(x, a)$ satisfies the soft Bellman expectation equation
\begin{equation}
    Q(x_t, a_t) = \E \left[ r_t + \alpha \E_{a_{t+1} \sim \pi(\cdot|x_{t+1})} \left[Q_{\text{target}}(x_{t+1}, a_{t+1}) - \gamma \log \pi(a_{t+1} | x_{t+1}) \right] \right]
\end{equation}
Note that the target value includes not only the discounted expected future soft Q-value but also the expected entropy term of the policy at the next state $x_{t+1}$.
The soft state-value function $V(x)$ is related by:
\begin{equation}
    V(x) = \E_{a \sim \pi(\cdot|x)} [ Q(x, a) - \gamma \log \pi(a | x) ]
\end{equation}
Substituting $V(x)$ into the Bellman equation for $Q(x, a)$ gives:
\begin{equation}
    Q(x_t, a_t) = \E \left[ r_t + \alpha V(x_{t+1}) \right]
\end{equation}
These relationships form the basis for the critic updates in SAC.

\subsubsection{Reparameterisation Trick}
SAC incorporates several crucial techniques:
\begin{itemize}
    \item \textbf{Reparameterisation Trick:} To obtain policy gradients, SAC uses the reparameterisation trick.
    Instead of sampling an action $a$ directly from the policy output distribution $\pi_\theta(\cdot|x)$, the algorithm samples a noise vector $\epsilon$ from a simple base distribution (e.g., standard Gaussian $\mathcal{N}(0, I)$) and then computes the action as a deterministic function $a = f_\theta(\epsilon; x)$.
    For a Gaussian policy, the actor network outputs the mean and standard deviation, i.e., $(\mu_\theta(x), \sigma_\theta(x)) = \pi_\theta(x)$, and the action is computed as $f_\theta(\epsilon; x) = \mu_\theta(x) + \sigma_\theta(x) \odot \epsilon$ where $\odot$ denotes element-wise multiplication.
    This makes the stochasticity external to the network parameters $\theta$, allowing gradients to flow back through the function $f_\theta$.
    \item \textbf{Automatic Temperature Tuning $\gamma$:} The temperature parameter $\gamma$ critically balances the reward maximisation and entropy maximisation objectives.
    Manually tuning $\gamma$ can be difficult and task-dependent.
    In \cite{haarnoja2018softb}, the authors modified the objective \eqref{eq:sac_objective} to a constrained optimisation problem where the goal is to maximise expected return subject to a constraint on the minimum expected policy entropy.
    The dual of this problem leads to an objective function for $\gamma$
    \begin{equation} \label{eq:sac_temperature}
        J(\gamma) = \E_{x,a} [ -\gamma (\log \pi_\theta(a | x) + \mathcal{H}_{\text{min}}) ]
    \end{equation}
    where $\mathcal{H}_{\text{min}}$ is the desired minimum expected entropy value
    $\gamma$ is then treated as a learnable parameter and updated via gradient descent on $J(\gamma)$.
    The automatic tuning of $\gamma$ is a significant improvement over the algorithm originally introduced in \cite{haarnoja2018soft} where $\gamma$ required manual tuning implicitly through reward scaling.
\end{itemize}

\subsubsection{Loss Functions and Optimisation}
Updates are performed using mini-batches sampled from the replay buffer:
\begin{enumerate}
    \item \textbf{Critic Loss:} The parameters $\phi_1$ and $\phi_2$ of the two Q-networks are updated to minimise the soft Bellman residual.
    The target value $Q_{\text{target}}$ uses the minimum of the two target Q-networks to mitigate overestimation bias:
    \begin{equation}
        Q_{\text{target}}(x_t,a_t) = r_t + \alpha \min_{i=1,2} Q_{\phi'_i}(x_{t+1}, a_{t+1}) - \gamma \log \pi_\theta(a_{t+1} | x_{t+1})
    \end{equation}
    The loss for each Q-network is then the mean squared error (MSE) between the predicted Q-value and the target value:
    \begin{equation}
        L_Q(\phi_i) = \frac{1}{N_B} \sum_{j=1}^{N_B} \left( Q_{\text{target}}^{(j)} - Q_{\phi_i}(x_t^{(j)}, a_t^{(j)}) \right)^2 \quad \text{for } i=1, 2
    \end{equation}
    \item \textbf{Actor Loss:} The actor parameters $\theta$ are updated to maximise the expected soft value.
    Using the reparameterisation trick ($a = f_\theta(\epsilon; x)$) and typically using the first Q-network for the gradient signal (though averaging or using the minimum is also possible), the loss is calculated as
    \begin{equation}
        L_\pi = -\frac{1}{N_B} \sum_{j=1}^{N_B} Q_{\phi_1}(x_t^{(j)}, f_\theta(\epsilon; x_t^{(j)})) + \gamma \log \pi_\theta(f_\theta(\epsilon; x_t^{(j)}) | x_t^{(j)})
    \end{equation}

    \item \textbf{Temperature Loss (optional):} The temperature parameter $\gamma$ is updated to minimise its objective function in \eqref{eq:sac_temperature}.
    Alternatively, the temperature can be fixed to a constant value.

    \item \textbf{Target Network Updates:} The target Q-network weights $\phi'_1, \phi'_2$ are updated using soft updates as in \eqref{eq:ddpg_target_network_update} or hard updates where the target networks are copied directly after a certain number of updates to the main parameters.
\end{enumerate}

\section {Experiments}

\subsection{Low Initial Wealth}

We consider experiments with low initial wealth of 1,000 which results in low market impact.
The parameters for the GBM, market impact and algorithms are provided in Appendix~\ref{app:low}.
In this setting, the market impact is low and we expect the optimal policy to be close that described in Section~\ref{sec:GBMbaseline}.
There are 4 key findings in this set of experiments:

\begin{enumerate}
  \item The use of GAE, introduced in Section~\ref{sec:ppo_gae}, to tune the right trade-off between bias and variance has a strong influence on the result.
  \item The clipping function of PPO reduces the likelihood of deviating from the optimal policy once it is found.
  \item Off-policy algorithms were unable to learn a good policy due to an inability to learn the right $Q$-function, which is likely caused by noise from the rewards.
  \item It takes at least 2m steps with the best algorithm (PPO) to converge to a good policy
\end{enumerate}

\subsubsection{Generalised Advantage Estimation} \label{sec:GAE}

The inherent stochasticity of the rewards will cause higher variance in the policy gradient.
GAE allows tuning of the bias variance trade-off by exponential weighting ($\lambda$) of multi-step advantages.

\begin{table}[h]
    \centering
    \begin{tabular}{cccccc}
    \toprule
    {} & \multicolumn{2}{c}{Agent growth rate} & Bankruptcies & \multicolumn{2}{c}{Baseline growth rate} \\
    \midrule
    {}$\lambda$ &       Mean &   MAD &                   Mean &                 Mean &   MAD \\
    \midrule
    0.00       &      -0.388 & 0.494 &                    1.3 &                0.115 & 0.001 \\
    0.25       &      -0.062 & 0.128 &                    0.1 &                0.113 & 0.004 \\
    0.50       &       0.052 & 0.033 &                    0.0 &                0.114 & 0.004 \\
    0.75       &       0.081 & 0.023 &                    0.0 &                0.112 & 0.002 \\
    0.80       &       0.100 & 0.015 &                    0.0 &                0.115 & 0.005 \\
    0.85       &       0.100 & 0.011 &                    0.0 &                0.112 & 0.004 \\
    0.90       &       0.100 & 0.009 &                    0.0 &                0.115 & 0.007 \\
    0.95       &       0.082 & 0.017 &                    0.0 &                0.113 & 0.004 \\
    1.00       &       0.079 & 0.014 &                    0.0 &                0.112 & 0.004 \\
    \bottomrule
    \end{tabular}
    \caption[Evaluation of PPO with different values of $\lambda$ for GAE]{Evaluation of trained PPO agent over 10 runs with different values of $\lambda$ for GAE. The agent growth rate is the mean of the growth rates of the agent's portfolio. MAD is the mean absolute deviation of the growth rates. The bankruptcies column shows the number of runs that resulted in a bankruptcy, which is when the wealth is at or below zero. The baseline growth rate is the mean of the growth rates of the policy described in Section~\ref{sec:OptimalPolicy}.}
    \label{tab:ppogae}
\end{table}

Table~\ref{tab:ppogae} shows the results from evaluating a trained PPO agent.
The objective is the log of final wealth divided by the initial wealth.
We divided this by the time horizon, which is equivalent to a per unit time growth rate shown as the mean growth rate.
The baseline policy consistently produces results close to the theorectical optimal of 0.114.
Any episode that results in bankruptcy is removed from the calculation as the growth rate is undefined.
The number of bankruptcies, which is when the wealth is at or below zero, and the mean absolute deviation (MAD) of the growth rate is also shown.

When a single step advantage ($\lambda=0$) is used, the agent is unable to learn a good policy.
This is likely to due to high bias, compounded by noisy rewards.
When the advantage is a Monte Carlo estimate ($\lambda=1$), the agent is able to learn a good policy.
However, the performance is much stronger when $\lambda$ is between 0.8 and 0.9 which suggests an appropriate trade-off is required to manage the noisy rewards.

Even with the best value of $\lambda$, there remains a gap to the optimal growth rate.
This is likely due to 3 factors:
\begin{enumerate}
  \item The use of stochastic optimisation which is inherently noisy.
  \item The added noise from the rewards.
  \item The flatness of the optimisation landscape near the optimal growth rate.
\end{enumerate}
The last point is partly visualised in a simplier setting of two stocks in Figure~\ref{fig:GT}.

In the implementation of A2C in the Stable Baselines 3 library \cite{raffin2021stable}, there is also the option to use GAE.
While A2C was also able to learn a good policy, the results were not as strong as PPO.
Results for A2C are in Appendix~\ref{app:a2c}.

\subsubsection{Clipping Function of PPO}

\begin{figure} [htbp!]
    \centering
    \subfigure[Weights with clipping]{
        \includegraphics[scale=0.2]{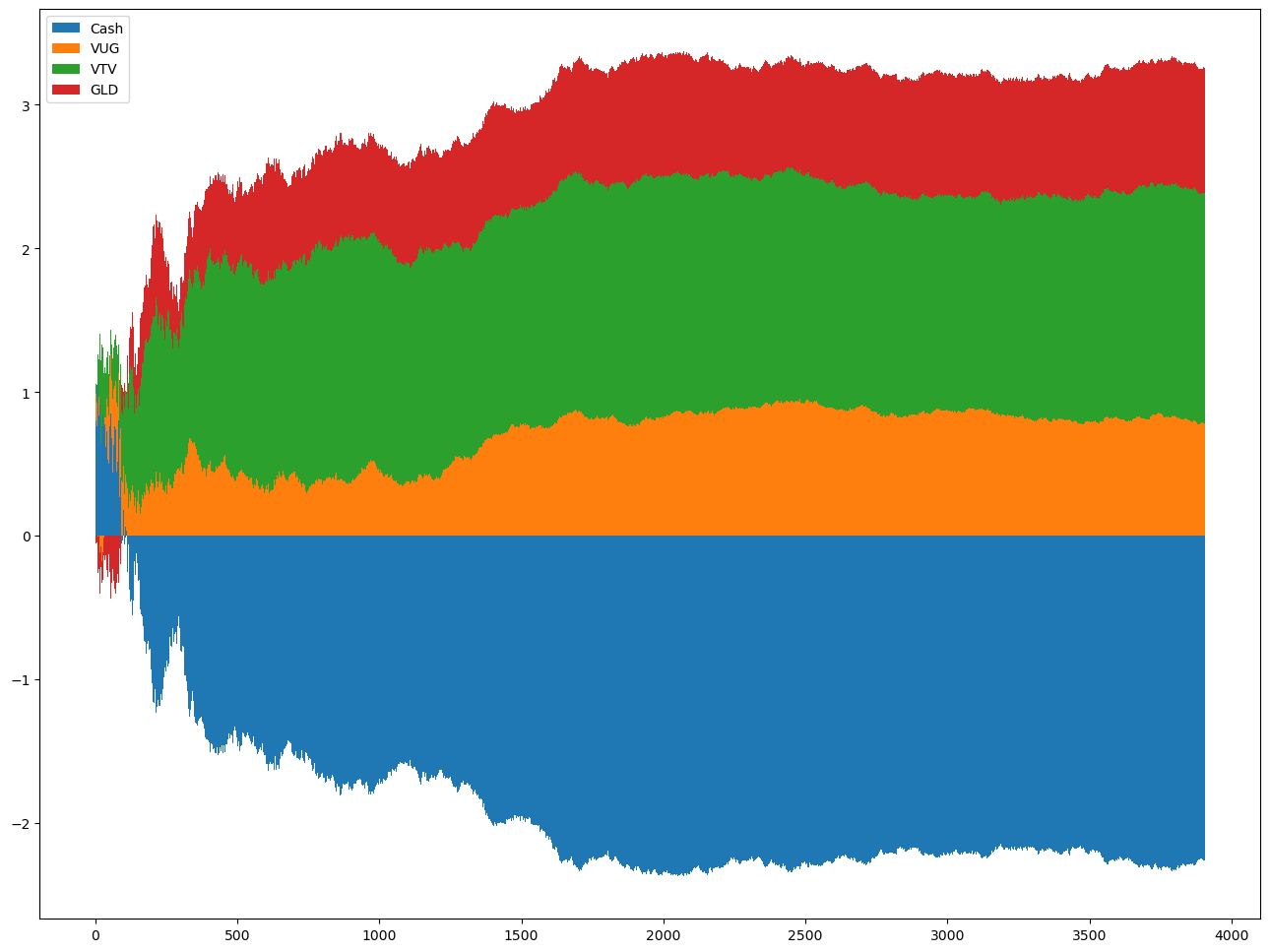}
        \label{fig:ppo_clip_weights}
    }
    \subfigure[MAD of weights in \ref{fig:ppo_clip_weights}]{
        \includegraphics[scale=0.2]{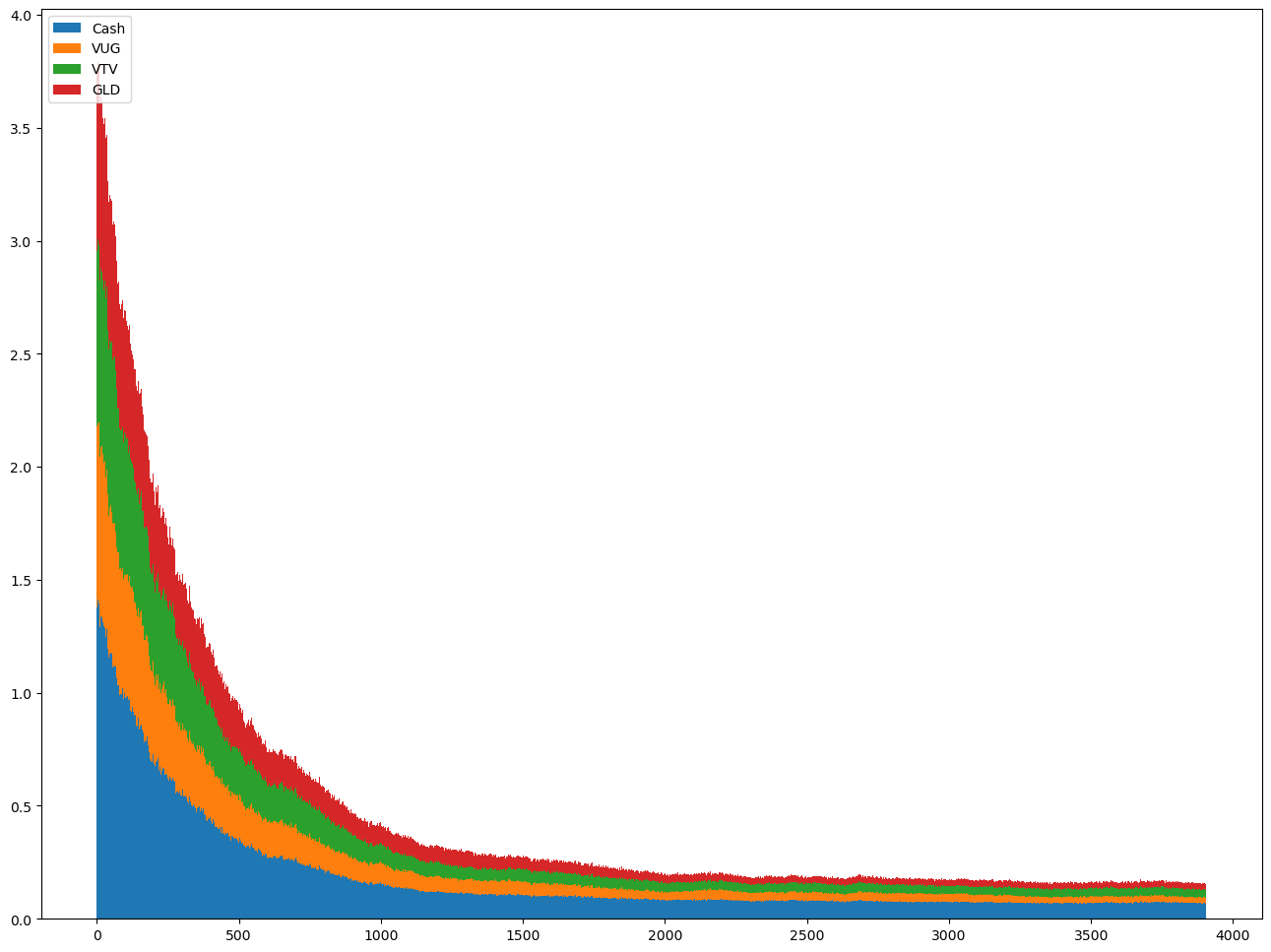}
        \label{fig:ppo_clip_mae}
    }
    \subfigure[Weights without clipping]{
        \includegraphics[scale=0.2]{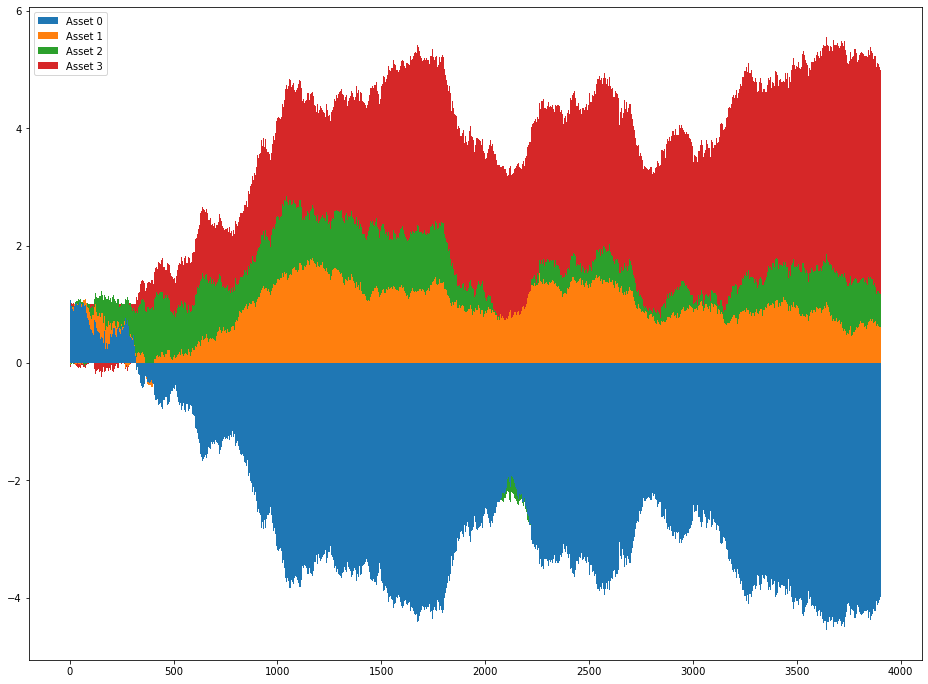}
        \label{fig:ppo_noclip_weights}
    }
    \subfigure[MAD of weights in \ref{fig:ppo_noclip_weights}]{
        \includegraphics[scale=0.2]{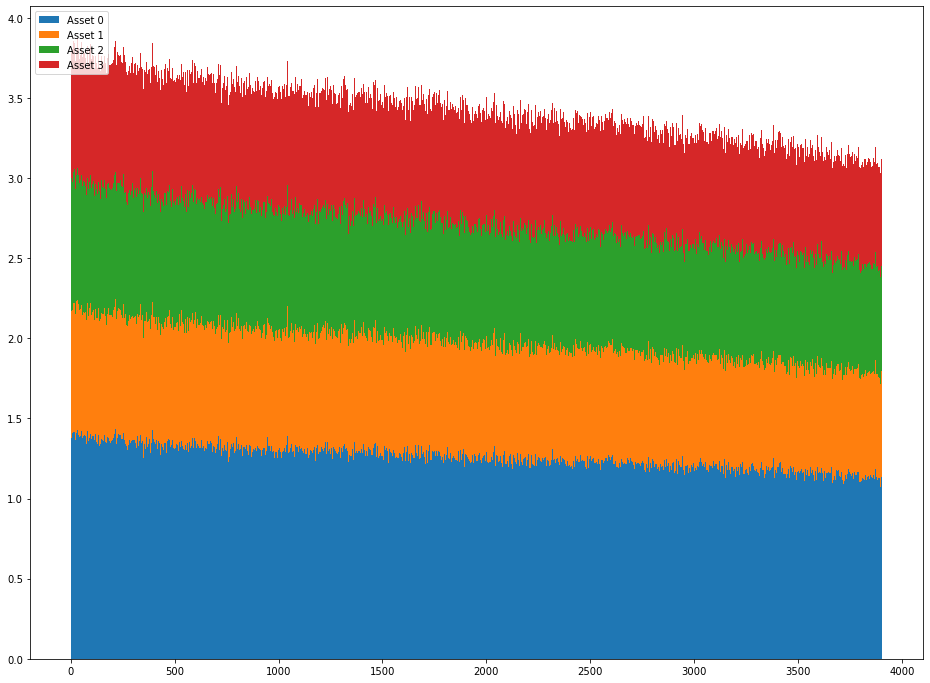}
        \label{fig:ppo_noclip_mae}
    }
    \subfigure[Weights with A2C]{
        \includegraphics[scale=0.2]{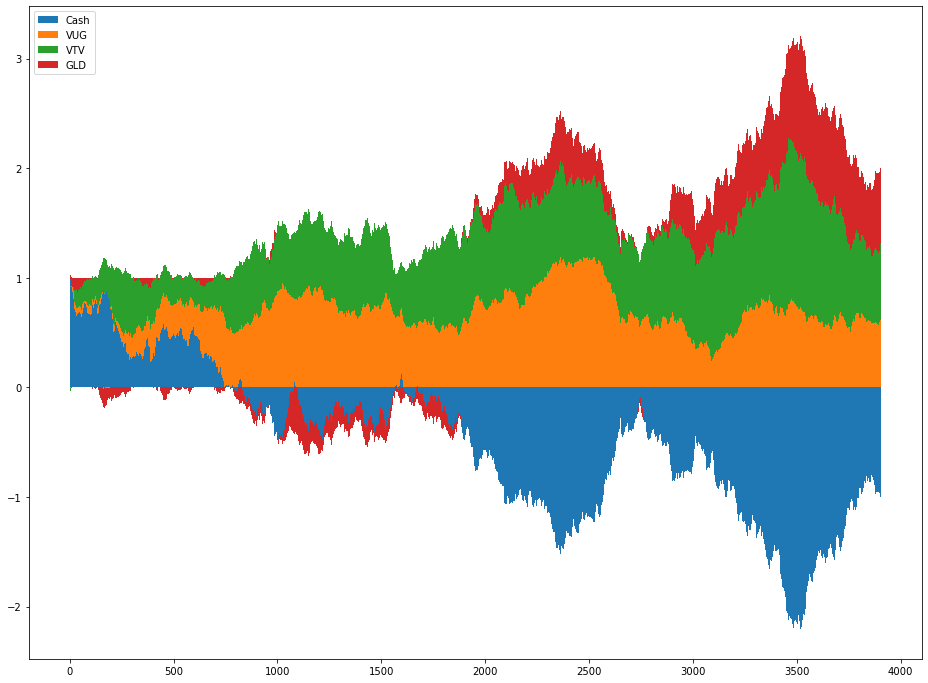}
        \label{fig:a2c_weights}
    }
    \subfigure[MAD of weights in \ref{fig:a2c_weights}]{
        \includegraphics[scale=0.2]{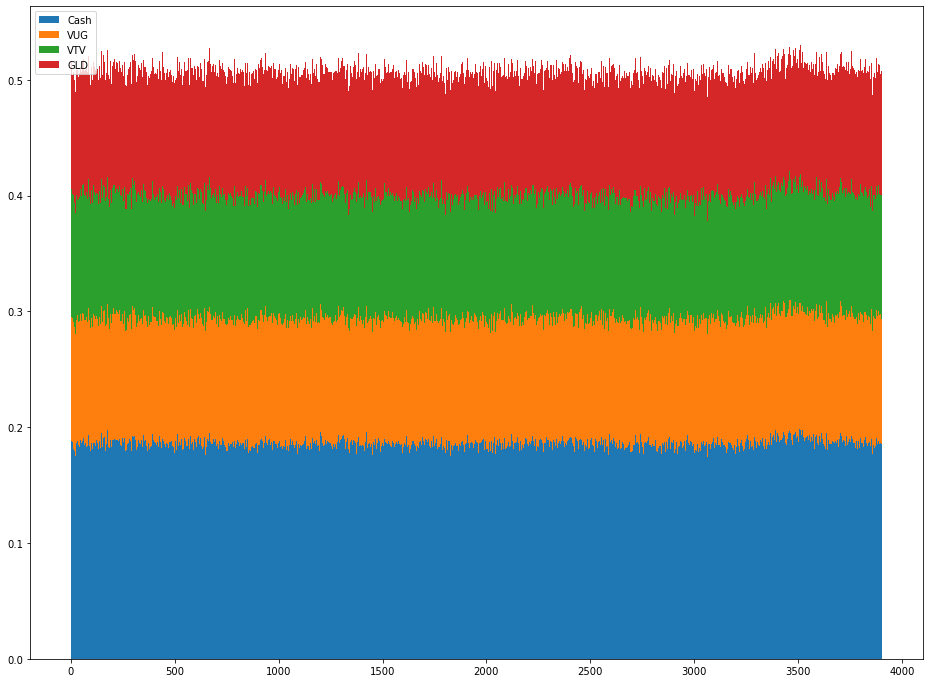}
        \label{fig:a2c_mae}
    }
    \caption[Importance of the clipping function in PPO]{The figures \ref{fig:ppo_clip_weights}, \ref{fig:ppo_noclip_weights} and \ref{fig:a2c_weights} show the evolution of the average portfolio weights (cash and 3 stocks) in an episode from a sample training run of PPO with clipping, PPO without clipping and A2C respectively. The figures \ref{fig:ppo_clip_mae}, \ref{fig:ppo_noclip_mae} and \ref{fig:a2c_mae} show the corresponding evolution of the mean absolute deviation (MAD) of the weights within an episode.}
    \label{fig:clip}
\end{figure}

The clipping function on the probability ratios of the new over old policy is a key component of PPO, as introduced in Section~\ref{sec:ppo_surrogate_objective}.
In the current setting, it was found to be important in preventing deviations from the optimal policy once it was found.
Figure~\ref{fig:ppo_clip_weights} is a stacked bar chart showing the evolution of \textbf{average portfolio weights in an episode} from a sample training run of PPO with clipping.
Each point on the x-axis represents an episode and the y-axis represents the value of the portfolio weight.
The baseline optimal weights are $[-1.72, 0.76, 0.66, 1.31]$ for cash (blue), VUG (orange), VTV (green) and GLD (red) respectively.
The key observation is that the weights stablises typically after 2,000 episodes of training.
The clipped PPO agent successful learns that the optimal policy has fixed weights as confirmed by Figure~\ref{fig:ppo_clip_mae},
  which shows the corresponding evolution of the \textbf{MAD of the weights within an episode}.

The contrast when turning off the clipping or with A2C can be seen in Figures~\ref{fig:ppo_noclip_weights} and \ref{fig:a2c_weights} respectively.
A2C is unable to converge towards a fixed weight policy asymptotically as seen in the unchanging MAD of weights in \ref{fig:a2c_mae}.
Note that one difference with A2C was the initial log standard deviation of the Gaussian policy had to be set much lower for the algorithm to converge.
For PPO without clipping, there was a trend towards lower MAD in the weights but the speed of convergence was orders of magnitude slower as seen in \ref{fig:ppo_noclip_mae}.
Note that without clipping, it was necessary to train only a single epoch per update to prevent instability.

\subsubsection{Off Policy Algorithms with Noisy Rewards}

\begin{figure} [htbp!]
    \centering
    \subfigure[Ground truth]{
        \includegraphics[scale=0.2]{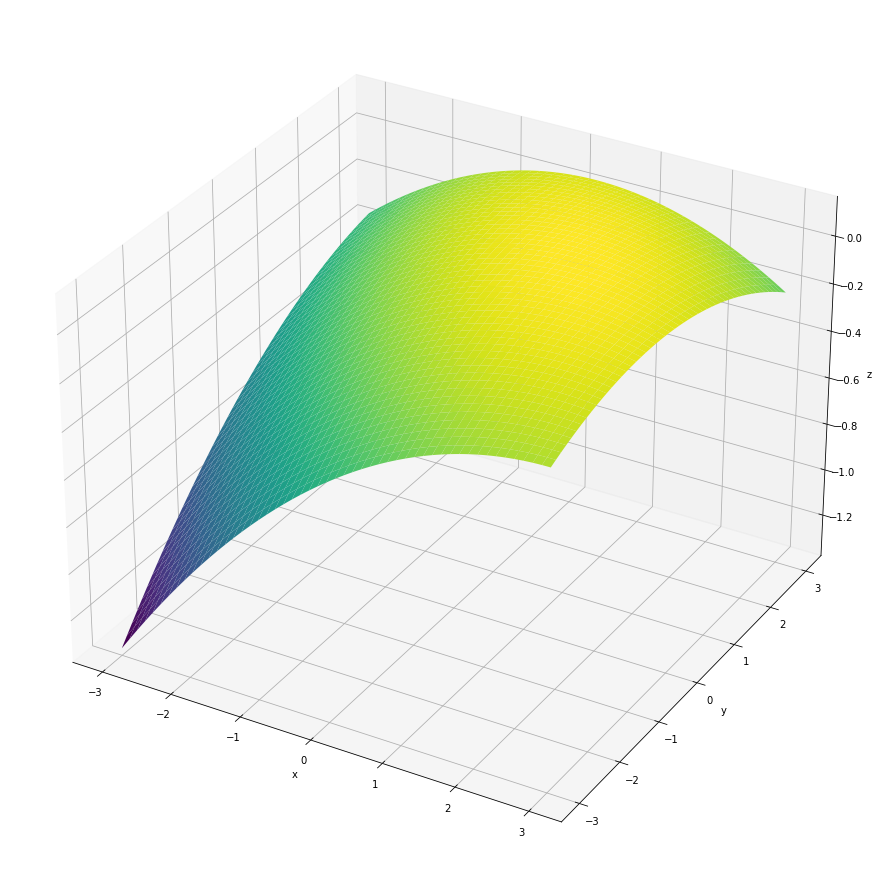}
        \label{fig:GT}
    }
    \subfigure[DDPG with noisy rewards]{
        \includegraphics[scale=0.2]{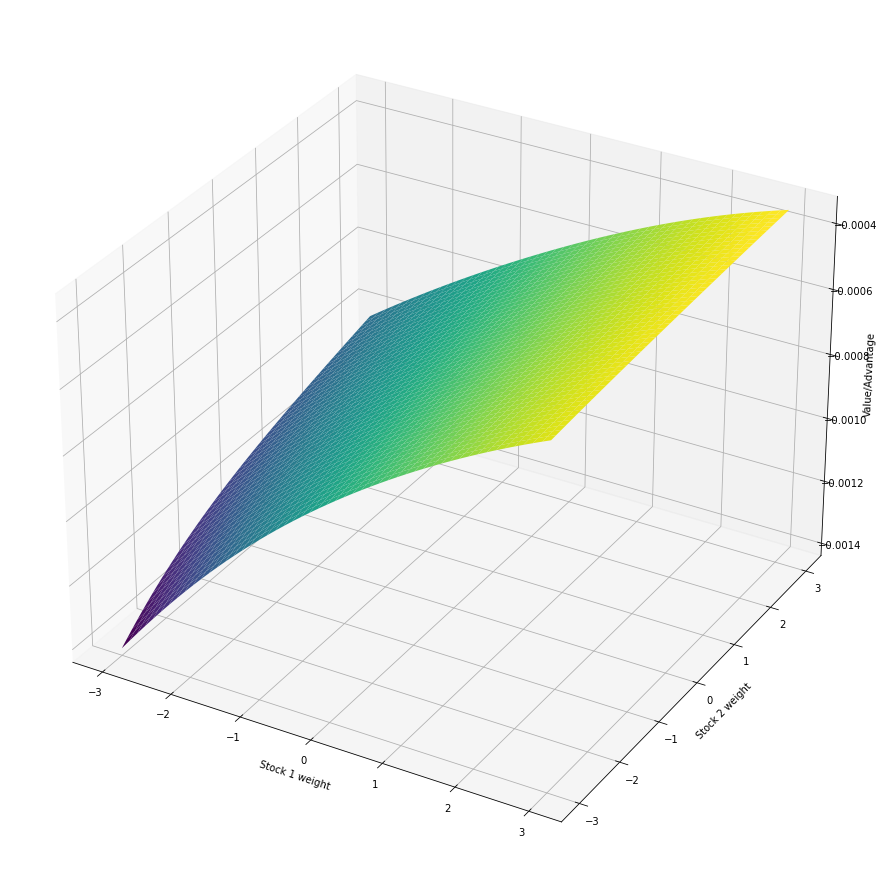}
        \label{fig:DDPG}
    }
    \subfigure[DDPG with noise free rewards]{
        \includegraphics[scale=0.2]{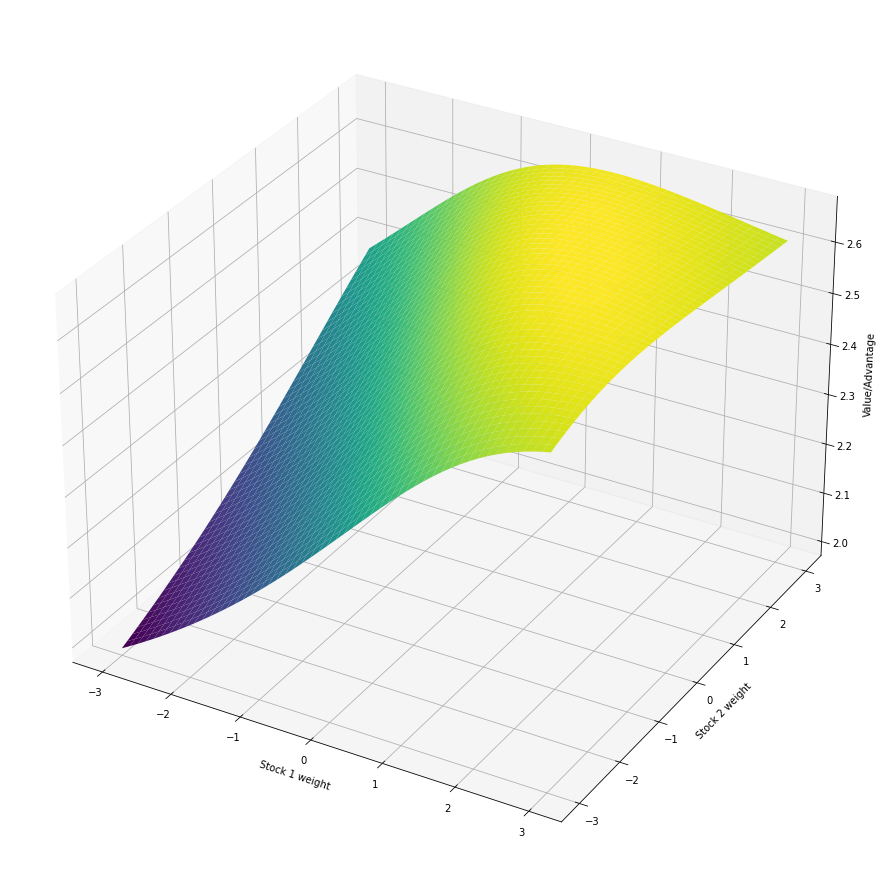}
        \label{fig:DDPG-GT}
    }
    \subfigure[DDPG using $\beta$-NLL loss on noisy rewards]{
        \includegraphics[scale=0.8]{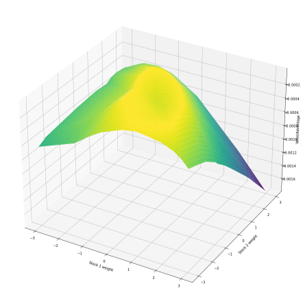}
        \label{fig:NLL}
    }
    \caption[Learning the $Q$-function]{The figures \ref{fig:GT}, \ref{fig:DDPG} and \ref{fig:DDPG-GT} show the ground truth $Q$-function, the $Q$-function learned by DDPG with noisy rewards and the $Q$-function learned by DDPG with noise free rewards respectively. The figure \ref{fig:NLL} shows the critic network trained using $\beta$-NLL loss with noisy rewards.}
    \label{fig:Q}
\end{figure}

The failure of the off-policy algorithms can be attributed to the noisy rewards received by the agent as we demostrate in this section.
A simplified experiment is conducted to investigate the difficulty of learning the correct $Q$-function in this noisy setting.
We reduce the action space to two stocks to visualise the $Q$-function.
Figure~\ref{fig:GT} shows the ground truth for the $Q$-function \textbf{assuming zero market impact}, which is independent of the state.
When the wealth of the agent is low, we would expect the $Q$-function in the environment with market impact to be similar.

Figure~\ref{fig:DDPG} shows the $Q$-function learned by the DDPG agent with low wealth after more than 10m steps.
The inability to learn the correct $Q$-function can be attributed to noise.
Figure~\ref{fig:DDPG-GT} shows the DDPG critic when the noisy rewards were replaced with expected rewards, i.e., noise free rewards.
The algorithm successfully converges to a good policy when the rewards are noise free.
Attempts to remedy this, such as increasing the critic network's size or using a critic with a $\beta$-NLL loss \cite{seitzer2022pitfalls} to learn a probabilistic $Q$-function, suited to the Gaussian noise, as shown in Figure~\ref{fig:NLL} were unsuccessful.

\subsubsection{Sample Efficiency}

As PPO was found to be the best performing algorithm, we investigate the sample efficiency of the algorithm.
Using tuned hyperparameters, we evaluate the growth rates of trained policies after increasing amount of training time steps.
The results are shown in Table~\ref{tab:sample_efficiency}.

\begin{table} [htbp!]
    \centering
    \begin{tabular}{lrrr}
    \toprule
    {} & \multicolumn{2}{l}{Agent growth rate} & Bankruptcies \\
    \midrule
    Time steps &        Mean &   MAD &                   Mean \\
    \midrule
    100k       &      -0.133 & 0.013 &                    0.0 \\
    500k       &       0.010 & 0.013 &                    0.0 \\
    1m         &       0.065 & 0.019 &                    0.0 \\
    2m         &       0.090 & 0.008 &                    0.0 \\
    3m         &       0.091 & 0.010 &                    0.0 \\
    4m         &       0.104 & 0.005 &                    0.0 \\
    5m         &       0.100 & 0.009 &                    0.0 \\
    \bottomrule
    \end{tabular}
    \caption[Sample efficiency of the algorithms]{Sample efficiency of PPO with low initial wealth. The time steps is the number of training steps. The other metrics are the same as in Table~\ref{tab:ppogae}.}
    \label{tab:sample_efficiency}
\end{table}

PPO takes at least 2m time steps to converge to a sufficiently well performing policy.
Note that if we take each period to represent a trading day, then 2m time steps corresponds to almost 8,000 years of training.
This is far too inefficient considering the amount of data that is typically available in financial markets.

\subsection{High Initial Wealth} \label{sec:highwealth}

When we increase the initial wealth of the agent, the effect of market impact becomes more pronounced.
The baseline optimal policy derived in Section~\ref{sec:MIbaseline} is able to attain similar performance as in the low wealth scenario , i.e., mean growth rate of 0.114.
Since PPO was the best performing algorithm in the low wealth scenario, we only compare the performance of PPO with the baseline policy in the high wealth scenario.

\begin{table} [htbp!]
    \centering
    \begin{tabular}{lrrr}
    \toprule
    {} & \multicolumn{2}{l}{Agent growth rate} & Bankruptcies \\
    \midrule
    Wealth &        Mean &   MAD &                   Mean \\
    \midrule
    100k &       0.085 & 0.010 &                    0.0 \\
    150k &       0.082 & 0.005 &                    0.0 \\
    200k &       0.061 & 0.012 &                    0.0 \\
    250k &       0.043 & 0.032 &                    0.0 \\
    300k &       0.045 & 0.023 &                    0.5 \\
    \bottomrule
    \end{tabular}
    \caption[Performance of PPO with different levels of initial wealth]{Performance of PPO with different levels of initial wealth. The metrics are the same as in Table~\ref{tab:ppogae}.}
    \label{tab:wealth}
\end{table}

It was found that PPO was converging towards policies that produces weights approaching fractions of the optimal weights akin to a fractional Kelly strategy \cite{davis2011fractional}.
This lowers the cost of market impact but results in lower growth rates as seen in Table~\ref{tab:wealth}.
Attempts to have the agent learn the gradual adjustment behaviour such as with larger networks, use of a recurrent neural network architecture and use of CRL \cite{hallak2015contextual} were unsuccessful.

\subsection{Regime Switching Model}

In the next set of experiments, we test whether DRL agents can adapt to the different regimes using contextual reinforcement learning (CRL) \cite{hallak2015contextual} and a hidden Markov model (HMM) \cite{rabiner1989tutorial} to learn to predict the current regime.

For the regime switching model, we use the dual regime parameters estimated from \cite{ang2002international} with a slight adjustment to prevent unrealistic levels of leverage as detailed in Appendix~\ref{app:rs}.
There are also 3 stocks in this model representing, US, Germany and UK markets.
The optimal weights, assuming no market impact, for each regime are as follows.

\begin{enumerate}
  \item \textbf{Bullish regime}: $[-1.72, 0.76, 0.66, 1.31]$ with an expected growth rate of 0.274
  \item \textbf{Bearish regime}: $[1.56, -2.18, 1.22, 0.40]$ with an expected growth rate of 0.104
\end{enumerate}
Based on the limiting distribution of the Markov chain, the expected growth rate of a policy that switches between these two sets of weights is 0.232 when there is no or low market impact.

When PPO is applied to this setting without modification, it is found to converge to the optimal \textbf{fixed weight policy}.
In other words, it finds a set of fixed weights that is used for both regimes with the highest expected growth rate.

\begin{figure} [htbp!]
    \centering
    \includegraphics[scale=0.6]{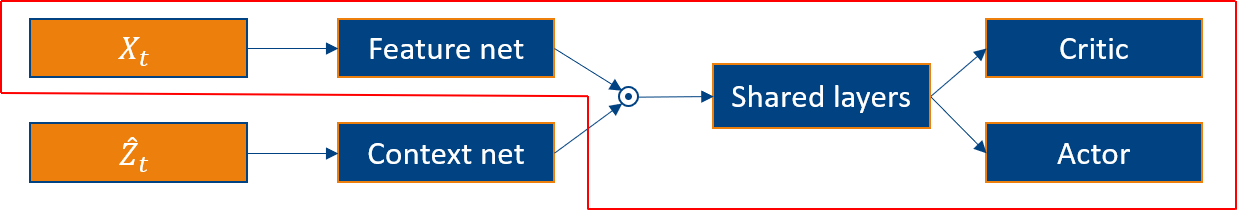}
    \caption{Network architecture for the regime switching model}
    \label{fig:context_net}
\end{figure}

A modification, as mentioned in Section~\ref{sec:RSM}, is required for the agent to learn a switching policy adapted to each regime.
A multivariate Gaussian HMM is trained on the historical log returns, derived from the state, to infer the regime index.
This serves as an additional context variable used in the network architecture shown in Figure~\ref{fig:context_net}.
The architecture used for the previous experiments is encased in the red box.
The additional context network takes in the learned regime, $\hat{Z}_t$, and outputs a vector of the same size as the output of the feature net.
These two are combined by element wise multiplication and fed into subsequent layers.
More details are in Appendix~\ref{app:rs}.

\subsubsection{Low Initial Wealth under Regime Switching Model}

\begin{figure} [htbp!]
    \centering
    \subfigure[Bullish regime weights]{
        \includegraphics[scale=0.2]{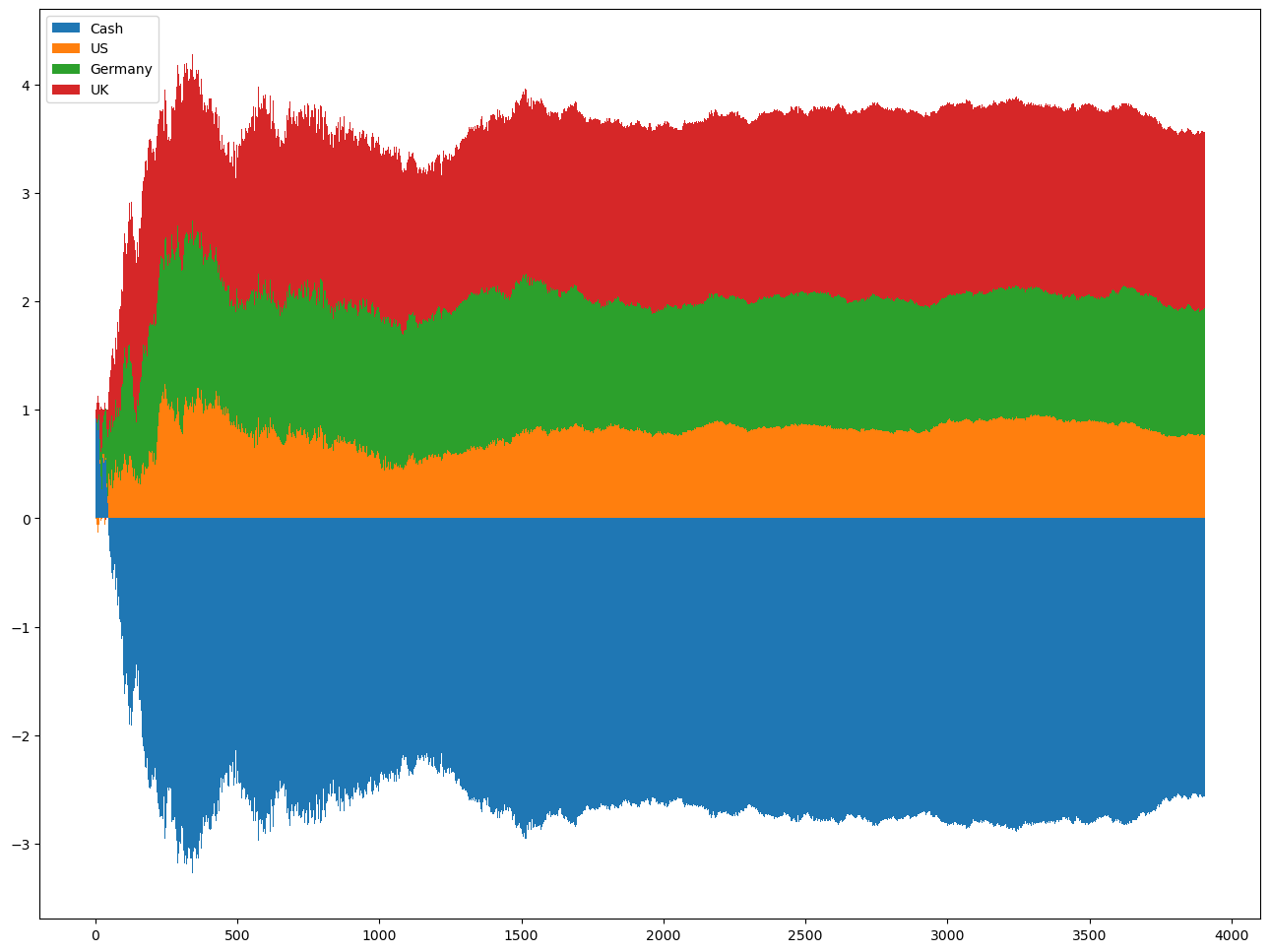}
        \label{fig:bull_weights}
    }
    \subfigure[MAD of weights in \ref{fig:bull_weights}]{
        \includegraphics[scale=0.2]{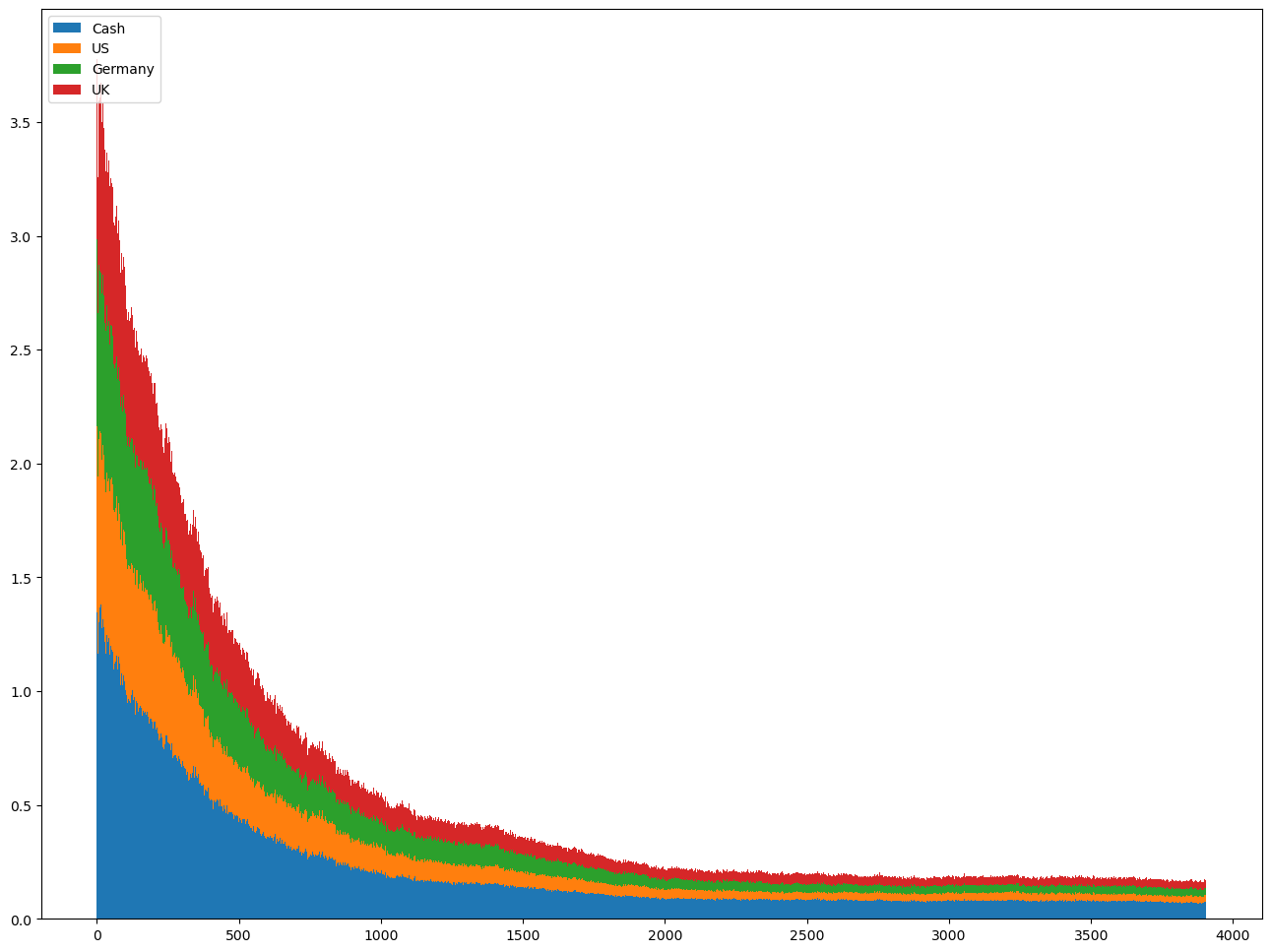}
        \label{fig:bull_mad}
    }
    \subfigure[Bearish regime weights]{
        \includegraphics[scale=0.2]{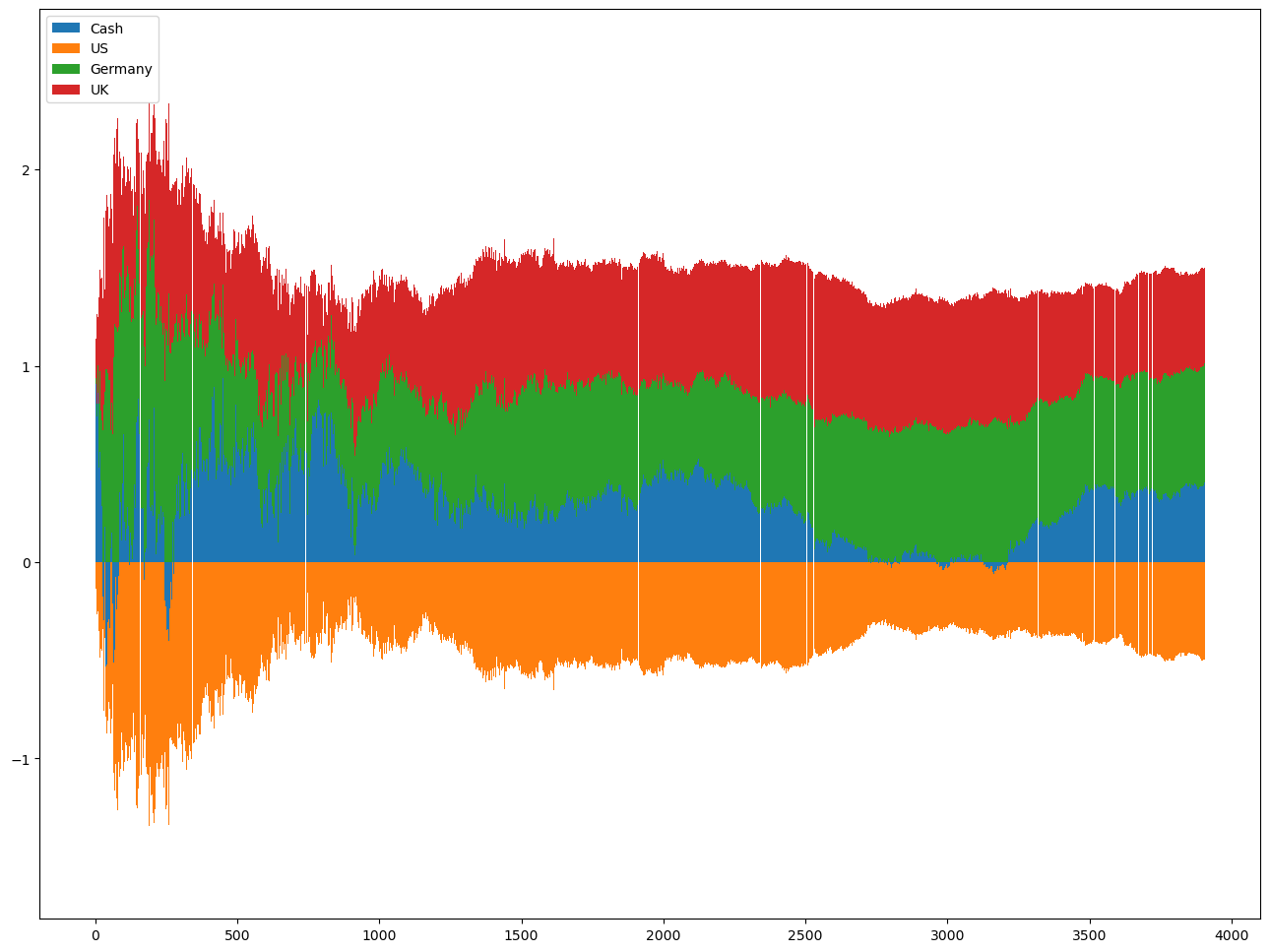}
        \label{fig:bear_weights}
    }
    \subfigure[MAD of weights in \ref{fig:bear_weights}]{
        \includegraphics[scale=0.2]{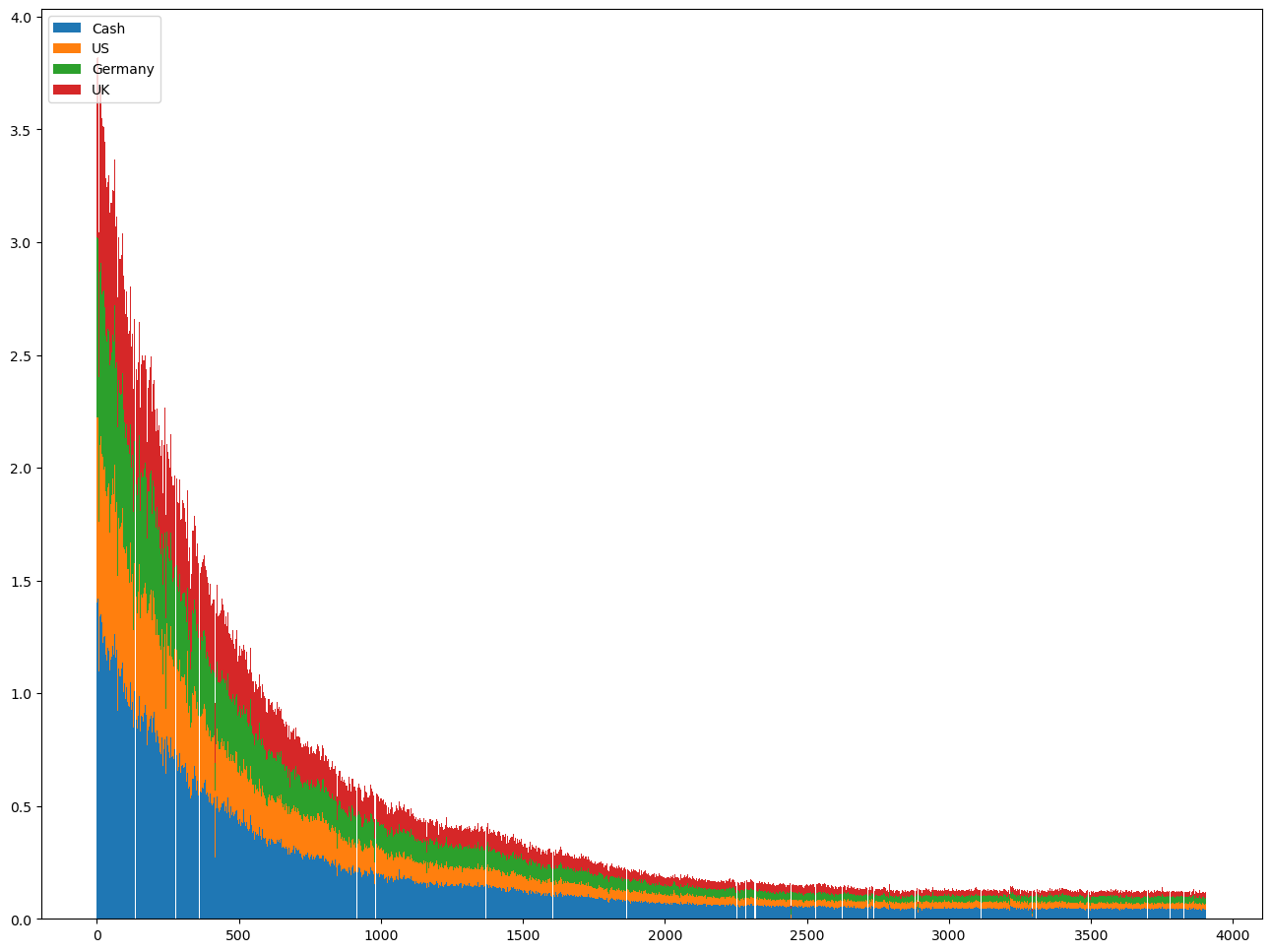}
        \label{fig:bear_mad}
    }
    \caption[PPO with context on the regime switching model with low wealth]{Training of PPO with context on the regime switching model with low wealth. The figures show the same type of charts as in Figure~\ref{fig:clip}.}
    \label{fig:hmm_low}
\end{figure}

We see in Figure~\ref{fig:hmm_low} the same type of charts as described for Figure~\ref{fig:clip}.
The key observation here is that the agent is now able to adapt different policies based on the predicted regime.
Figures~\ref{fig:bull_weights} and \ref{fig:bear_weights} show the weights of the stocks in the bullish and bearish regimes respectively.
In particular, the agent has learned to reduce exposure and also short the US market in the bearish regime, consistent with the optimal policy.
The agent also correctly learns that each regime's optimal policy is still a fixed weight policy.
Figures~\ref{fig:bull_mad} and \ref{fig:bear_mad} show that it is converging towards fixed weights.

Over 10 runs, the agent achieves a mean growth rate of 0.161 with a MAD of 0.019.
There are a few reasons that contribute to the wider gap from the optimal growth rate beyond those discussed in Section~\ref{sec:GAE}.
Firstly, a regime switch happens for at least one period before the HMM is able to predict the change and the agent is able to adapt compared to the foresight used in the baseline.
Although it is an unfair comparison, we use foresight in the baseline as it provides a more accurate check on correctness when compared to the theorectical optimal.
Secondly, while the HMM is able to predict the regime with a high accuracy of 0.971 on average, there is still an error rate which contributes to a dip in the growth rate.
Overall, we see that the agent is able to learn the right attributes of the optimal policy.

\subsubsection{High Initial Wealth under Regime Switching Model} \label{sec:rs_highwealth}

\begin{figure} [htbp!]
  \centering
  \subfigure[Policy with distinct weights for each regime]{
    \includegraphics[scale=0.2]{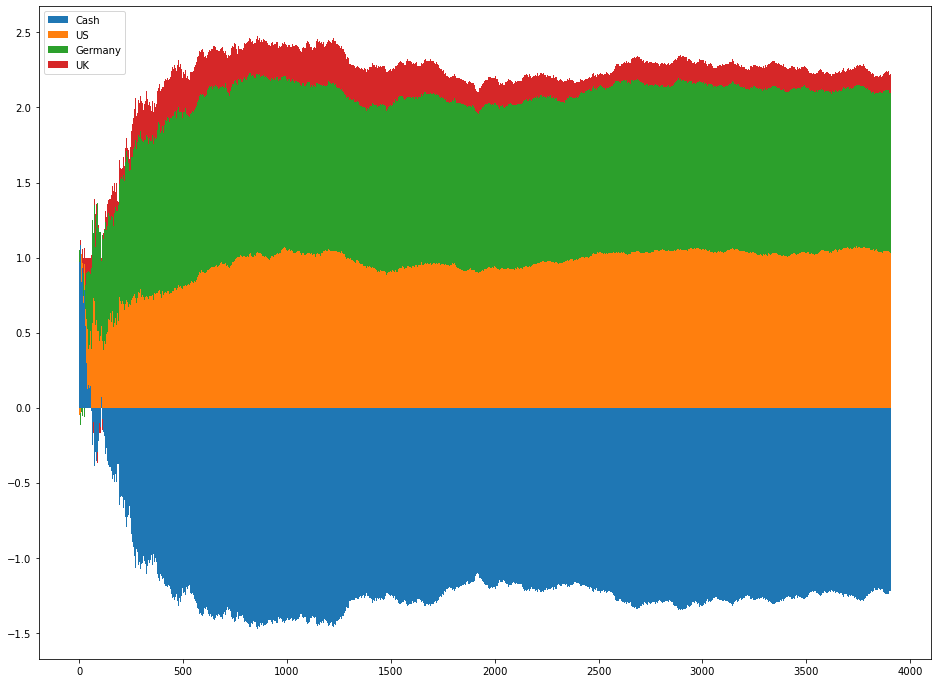}
    \includegraphics[scale=0.2]{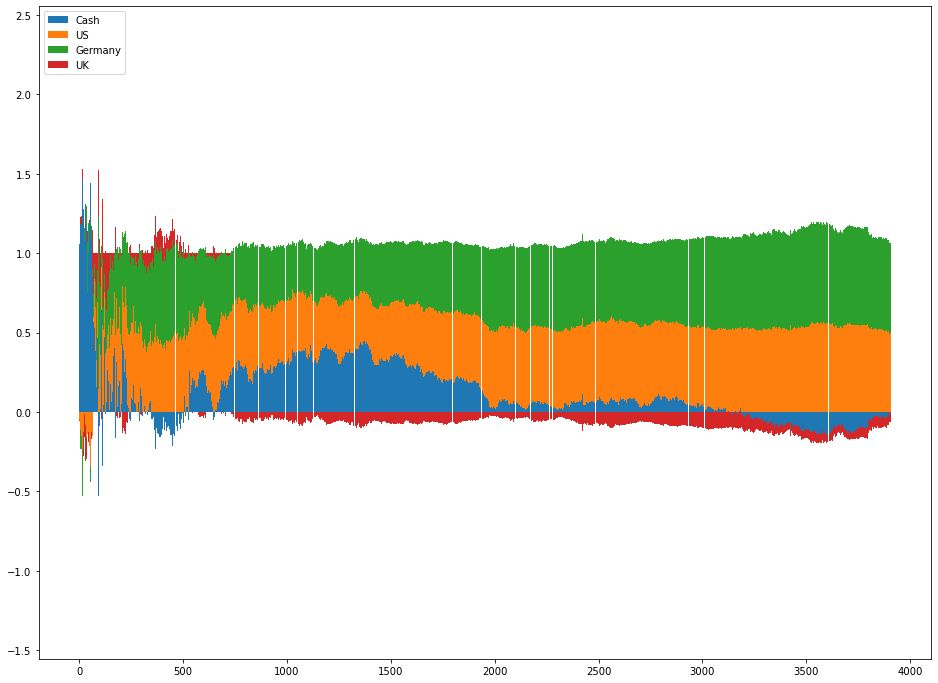}
    \label{fig:distinct}
  }
  \subfigure[Policy with similar weights for each regime]{
    \includegraphics[scale=0.2]{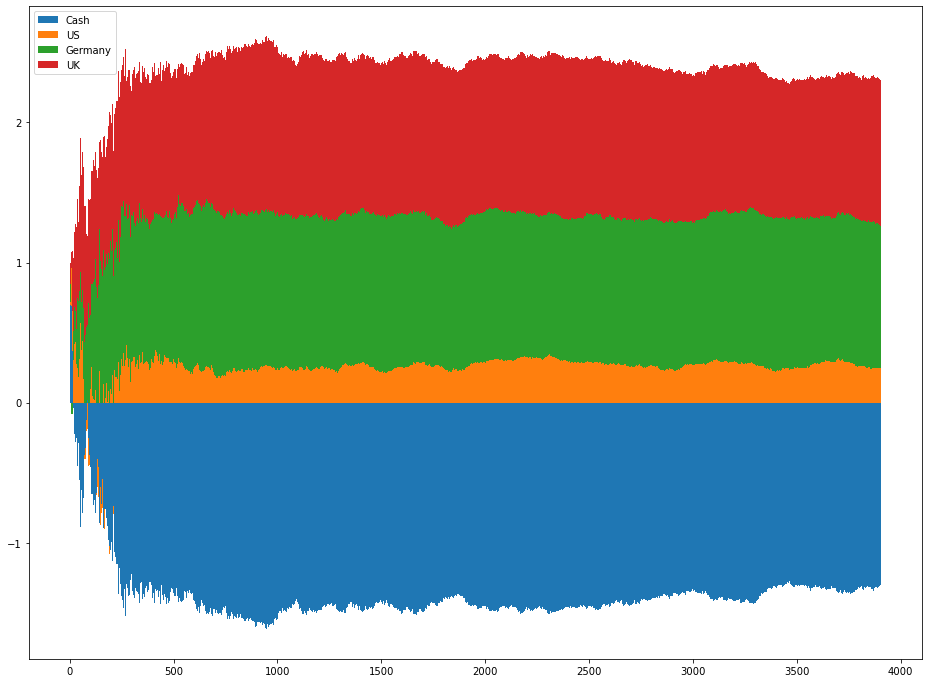}
    \includegraphics[scale=0.2]{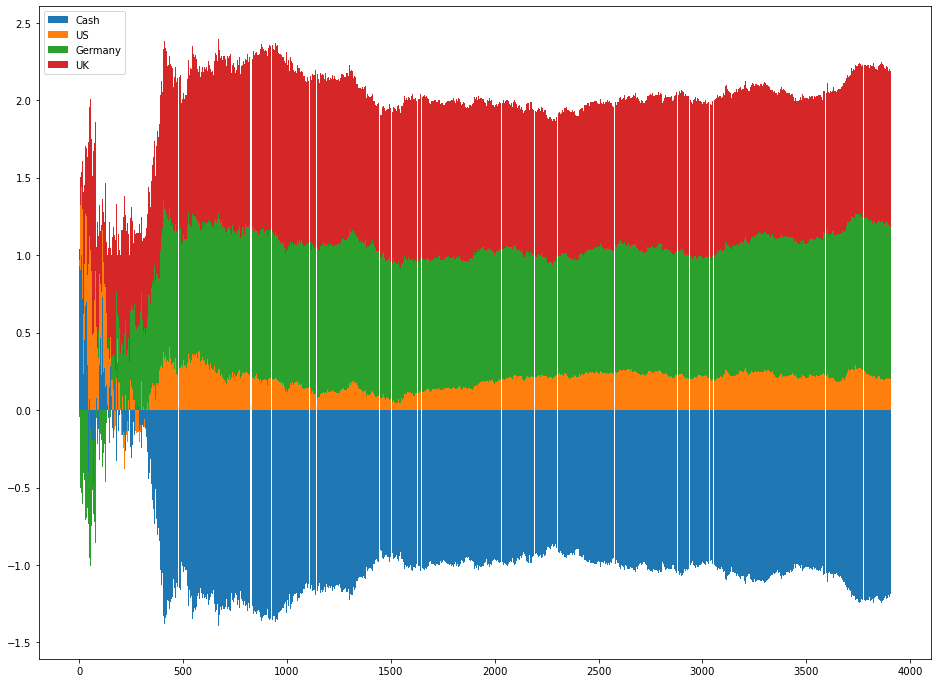}
    \label{fig:similar}
  }
  \caption{PPO with context on the regime switching model with high wealth}
  \label{fig:hmm_high}
\end{figure}

As discussed in Section~\ref{sec:RSbaseline}, the heuristically derived policy is used as the baseline.
Under the higher wealth conditions set for this set of experiments, it achieves a mean growth rate of 0.164 with MAD of 0.004.

The PPO+HMM agent's results are mixed across multiple runs with a mean growth rate of 0.088 and MAD of 0.016.
The common thread across all runs is that the agent seeks policies akin to the fractional Kelly strategy seen in Section~\ref{sec:highwealth} to limit the cost of market impact for each regime.
In some cases, the policy converges towards more distinct policies for each regime as shown in Figure~\ref{fig:distinct} which achieved a growth rate of 0.091
In others, the policy has similar weights for both regimes as shown in Figure~\ref{fig:similar} resulting in low costs and achieved a growth rate of 0.129.
Overall, there is higher variance in the resulting policies as higher market impact costs delivers more noise into the rewards.

\section{Conclusion} \label{sec:conclusion}

The common benckmark DRL algorithms were evaluated in the task of portfolio optimisation under a simulator based on geometric Brownian motion with the Bertsimas-Lo market impact model.
The optimal policy under no market impact can be derived analytically and serves as an upper bound when market impact is included.
This provides the baseline to evaluate the performances of the algorithms.

Due to the noisy nature of the problem, off-policy methods struggled to learn a good policy.
The on-policy methods used generalised advantage estimation to reduce the variance in the policy gradient estimator and successfully converged to a close to optimal policy.
Among the two on-policy methods, PPO's clipping function was found to be more robust to the noise in terms of not drifting away once it had converged.
In a more challenging and realistic setting of regime changes, we used a HMM to predict the regime and provide the appropriate context to the agent.
This allowed the PPO algorithm to adapt different policies to each regime.

In all the cases, the sample efficiency is a limiting factor in the application of DRL to the problem of portfolio optimisation.
With the best performing algorithm PPO and the simplest setting with low wealth and no regime changes, it took around 2m steps of training to learn a policy that is close to optimal.
If we translate this into the number of trading days, it represents almost 8,000 years of data points.
With financial data, we only see one realisation of the market with no option to reset the environment.
Therefore, the results suggests large improvements on sample efficiency is required for successful application of DRL to the problem.

It is therefore not surprising that another avenue of research has emergered, which is to explore the use of generative models trained on real data to generate realistic synthetic data \cite{buehler2020data, wiese2020quant, koshiyama2021generative,lu2024generative} which can be used to train DRL agents.
As these are the early days of DRL in finance, and in particular portfolio optimisation, it remains to be seen how the field will develop and address these issues.

\section{Acknowledgements}
We would like to thank Julian Sester for his valuable feedback on the paper.

\bibliographystyle{IEEEtran}
\bibliography{literature}

\newpage

\section*{Appendix}

\appendix

\section{Experimental Results for A2C} \label{app:a2c}

The same experimental setup as described in Section~\ref{app:low} was used to evaluate A2C.
The $\lambda$ parameter was varied and for each value, 10 runs were performed.
Results are shown in Table~\ref{tab:a2c_low}.

\begin{table} [htbp!]
    \centering
    \begin{tabular}{cccccc}
    \toprule
    {} & \multicolumn{2}{c}{agent growth rate} & bankruptcies & \multicolumn{2}{c}{baseline growth rate} \\
    \midrule
    {}$\lambda$ &       Mean &   MAD &                   Mean &                 Mean &   MAD \\
    \midrule
    0.00       &      -0.527 & 0.314 &                    0.3 &                0.114 & 0.005 \\
    0.25       &      -0.061 & 0.176 &                    0.2 &                0.111 & 0.004 \\
    0.50       &       0.023 & 0.075 &                    0.0 &                0.112 & 0.005 \\
    0.75       &       0.082 & 0.010 &                    0.0 &                0.115 & 0.004 \\
    0.90       &       0.089 & 0.008 &                    0.0 &                0.111 & 0.006 \\
    0.95       &       0.087 & 0.010 &                    0.0 &                0.116 & 0.006 \\
    1.00       &       0.065 & 0.029 &                    0.0 &                0.118 & 0.005 \\
    \bottomrule
    \end{tabular}
    \caption[Evaluation of A2C with different values of $\lambda$ for GAE]{Evaluation of trained A2C agent over 10 runs with different values of $\lambda$ for GAE. The agent growth rate is the mean of the growth rates of the agent's portfolio. MAD is the mean absolute deviation of the growth rates. The bankruptcies column shows the number of runs that resulted in a bankruptcy which is when the portfolio value goes below zero. The baseline growth rate is the mean of the growth rates of the policy described in Section~\ref{sec:OptimalPolicy}.}
    \label{tab:a2c_low}
\end{table}

\section{Common Parameters for all Experiments} \label{app:common}

The common parameters for all experiments are shown in Table~\ref{tab:common_params}.

\begin{table} [htbp!]
    \centering
    \begin{tabular}{lrrr}
    \toprule
    Variable & Value \\
    \midrule
    T (Investment horizon) & 5 \\
    Number of periods per unit of time & 256 \\
    Number of periods in an episode & 1280 \\
    $\eta$ (temporary impact factor) & 1e-9 \\
    $\gamma$ (permanent impact factor) & 1e-7 \\
    $l$ (historical window length) & 60 \\
    \bottomrule
    \end{tabular}
    \caption{Common parameters for all experiments.}
    \label{tab:common_params}
\end{table}

\section{Experiment Parameters for Non-regime Switching Model} \label{app:low}

The interest rate for cash was set to 0.04. The GBM parameters are shown in Table~\ref{tab:gbm_low}.

\begin{table} [htbp!]
    \centering
    \begin{tabular}{lrrr}
    \toprule
    Ticker & VUG & VTV & GLD \\
    \midrule
    $\mu$ (drift)         & 0.124 & 0.105 & 0.072 \\
    $\sigma$ (volatility) & 0.255 & 0.209 & 0.145 \\
    $\rho$ (correlation)  & 0.81/0.12 & 0.81/0.08 & 0.12/0.08 \\
    \bottomrule
    \end{tabular}
    \caption{Geometric Brownian motion parameters}
    \label{tab:gbm_low}
\end{table}

All experiments with non-regime switching model used 2 layers of 64 neurons in the feature extractor with a Tanh activation and 1 linear layer for the actor and critic.
For the off-policy algorithms, various hyperparameters were tested but none successfully produced a policy that could consistently approach the baseline.

The hyperparameters for PPO are shown in Table~\ref{tab:ppo_hyp} and the hyperparameters for A2C are shown in Table~\ref{tab:a2c_hyp}.

\begin{table} [htbp!]
    \centering
    \begin{tabular}{lrrr}
    \toprule
    Hyperparameter                & Value \\
    \midrule
    $\alpha$ (discount factor)     & 0.99 \\
    Learning rate                 & 0.0003 \\
    Batch size                    & 64 \\
    Num steps between updates     & 1280 \\
    Num epochs per update         & 10 \\
    Clip range                    & 0.2 \\
    GAE $\lambda$                 & 0.9 (unless otherwise specified) \\
    Initial log std               & 0 \\
    Max grad norm                 & 0.5 \\
    Value function loss coefficient & 1.0 \\
    Entropy loss coefficient        & 0 \\
    \bottomrule
    \end{tabular}
    \caption{Hyperparameters for PPO}
    \label{tab:ppo_hyp}
\end{table}

\begin{table} [htbp!]
    \centering
    \begin{tabular}{lrrr}
    \toprule
    Hyperparameter & Value \\
    \midrule
    $\alpha$ (discount factor) & 0.99 \\
    Learning rate & 0.0001 \\
    Batch size & 256 \\
    Num steps between updates & 256 \\
    Num gradient steps per update & 1 \\
    GAE $\lambda$ & 0.9 (unless otherwise specified) \\
    Initial log std & -2.0 \\
    Value function loss coefficient & 1.0 \\
    \bottomrule
    \end{tabular}
    \caption{Hyperparameters for A2C}
    \label{tab:a2c_hyp}
\end{table}

\section{Experiment Parameters for Regime Switching Model} \label{app:rs}

The interest rate for cash was set to 0.05 for the bull regime and 0.01 for the bear regime.
The GBM parameters for the regime switching model are shown in Table~\ref{tab:gbm_rs}.
The growth rate for the US market was adjusted down as the original level was higher than both the Germany and UK market.
Since it also had lower volatility than those markets, it resulted in optimal portfolio weights that were extremely skewed towards the US market with unrealistic leverage.

\begin{table} [htbp!]
    \centering
    \begin{tabular}{llrrr}
    \toprule
    Regime & Parameter & US & Germany & UK \\
    \midrule
    Bull & $\mu$ (drift)          & 0.103     & 0.138     & 0.140 \\
    Bull & $\sigma$ (volatility)  & 0.120     & 0.166     & 0.166 \\
    Bull & $\rho$ (correlation)   & 0.41/0.26 & 0.41/0.43 & 0.26/0.43 \\
    \midrule
    Bear & $\mu$ (drift)          & -0.021    & 0.097     & 0.042 \\
    Bear & $\sigma$ (volatility)  & 0.216     & 0.379     & 0.288 \\
    Bear & $\rho$ (correlation)   & 0.60/0.45 & 0.60/0.45 & 0.45/0.45 \\
    \bottomrule
    \end{tabular}
    \caption{Geometric Brownian motion parameters for the regime switching model}
    \label{tab:gbm_rs}
\end{table}

The per period transitional probabilities of the Markov chain are shown in Table~\ref{tab:rs_prob}.
The original probabilities were estimated from \cite{ang2002international} were estimated for monthly returns.
We rescale them using Equation \eqref{eq:ctmc_rescale} to daily returns which are closer to the setting of the experiment.
Specifically, we solve for the infinitesimal generator matrix $Q$ where $e^{\frac{1}{12}Q}$ equals the parameters in \cite{ang2002international}.
Then we get the required probabilities by computing $e^{\frac{1}{256}Q}$.

\begin{table} [htbp!]
    \centering
    \begin{tabular}{lrr}
    \toprule
    Regime & $\mathbb{P}$(Bull|Regime) & $\mathbb{P}$(Bear|Regime) \\
    \midrule
    Bull & 0.997 & 0.003 \\
    Bear & 0.009 & 0.991 \\
    \bottomrule
    \end{tabular}
    \caption{Transition probabilities of the Markov chain}
    \label{tab:rs_prob}
\end{table}

\begin{figure} [htbp!]
  \centering
  \includegraphics[scale=0.5]{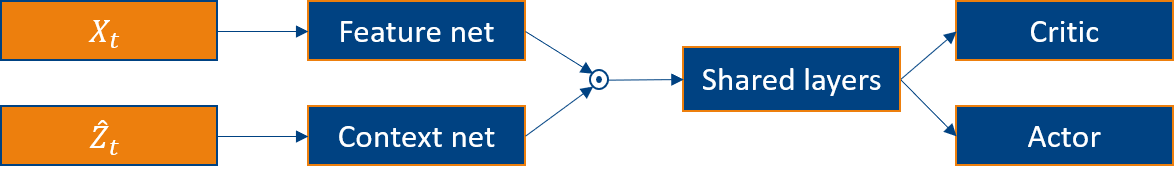}
  \caption{Network architecture for the regime switching model}
  \label{fig:context_net2}
\end{figure}

The structure of the neural network architecture is shown in Figure~\ref{fig:context_net2}.
The feature net has 3 layers with 256, 128 and 64 neurons respectively and a Tanh activation function.
The regime net has 3 layers with 64 neurons each and a ReLU activation function.
The shared layers consists of 2 layers with 64 neurons each and a Tanh activation function.
The actor and critic each have a single linear layer.

The multivariate Gaussian HMM was trained using the Viterbi algorithm \cite{rabiner1989tutorial} implemented by hmmlearn \url{https://github.com/hmmlearn/hmmlearn}.
Due to the possibility of no bear regime appearing in a single episode, the HMM is trained for the first 10 episodes of each experimental run.
Thereafter, the learned parameters are frozen.
The hyperparameters for the algorithm are shown in Table~\ref{tab:hmm}.
Hyperparameters not shown are using default values.

\begin{table} [htbp!]
    \centering
    \begin{tabular}{lrr}
    \toprule
    Hyperparamter & Value \\
    \midrule
    Number of components (regimes)                      & 2 \\
    Number of initialisations                           & 10 \\
    Max number of iterations per initialisation         & 100 \\
    Log likelihood convergence threshold                & 1e-7 \\
    Prior for mean                                      & 1e-4 \\
    Prior for covariance                                & 1e-4 \\
    Min covariance value                                & 1e-6 \\
    \bottomrule
    \end{tabular}
    \caption{Hyperparameters for the HMM}
    \label{tab:hmm}
\end{table}

\end{document}